\documentclass[twocolumn,preprintnumbers,amsmath,amssymb,longbibliography]{revtex4-2}

\usepackage{graphicx}
\usepackage{dcolumn}
\usepackage{float} 
\usepackage{bbm}

\usepackage{setspace}

\begin{document}

\title{ Polariton Localization and Dispersion Properties of  Disordered Quantum Emitters in Multimode  Microcavities } 

\author{ Georg Engelhardt$^{1,2,3}$    }

\author{ Jianshu Cao$^{4} $}
\email{jianshu@mit.edu}

\affiliation{%
	$^1$Shenzhen Institute for Quantum Science and Engineering, Southern University of Science and Technology, Shenzhen 518055, China\\
	$^2$International Quantum Academy, Shenzhen 518048, China\\
	$^3$ Guangdong Provincial Key Laboratory of Quantum Science and Engineering, Southern University of Science and Technology, Shenzhen, 518055, China.\\
	$^4$Department of Chemistry, Massachusetts Institute of Technology, 77 Massachusetts Avenue,
	Cambridge, Massachusetts 02139, USA
}

\date{\today}

\pacs{
  }

\begin{abstract}
	Experiments have demonstrated that the strong light-matter coupling in polaritonic microcavities significantly enhances transport. Motivated by these experiments, we have solved the disordered multimode Tavis-Cummings model in the thermodynamic limit and used this solution to analyze its dispersion and localization properties. The solution implies that wave-vector-resolved spectroscopic quantities can be described by single-mode models, but  spatially resolved quantities require the multimode solution.  Nondiagonal elements of the Green’s function decay exponentially with distance, which defines the coherence length. The coherence length is strongly correlated with the photon weight and exhibits inverse scaling with respect to the Rabi frequency and an unusual dependence on disorder. For energies away from the average molecular energy $E_{\text{M}}$ and above the confinement energy $E_C$, the coherence length rapidly diverges such that it exceeds the photon resonance wavelength $\lambda_0$.  The rapid divergence allows us to differentiate the localized and delocalized regimes and identify the transition from diffusive to ballistic transport. 
\end{abstract}

\maketitle

\allowdisplaybreaks

\textit{Introduction.}---
The spatial confinement of the light field in  microcavities gives rise to dispersive polaritons with outstanding spectroscopic properties~\cite{ Garcia-Vidal2021} and establishes an alternative channel for charge and energy transport different from  the short-range hopping. Recent experimental measurements of  microcavities have found that transport can be enhanced by  orders of magnitude~\cite{Lerario2017,Orgiu2015,Rozenman2018,Hou2020,Krainova2020,Balasubrahmaniyam2023}.   A thorough description is  challenging because of the large number of light modes in the cavity and  the energetic, spatial and orientational disorder.

Many theoretical models describe the light field by a single cavity mode,  which is coupled to a macroscopic number of quantum emitters~\cite{Chavez2021,Engelhardt2022,Feist2015,Sommer2021,Spano2015,Shammah2017,Botzung2020,Dubail2022,Zhang2021,Gera2022a,Cohn2022,Herrera2016,Houdre1996,Xiang2019,Reitz2018,Christian2019,Cao2022,Cui2022,FinkelsteinShapiro2023,Zhang2023}. Recent  investigations have predicted an intriguing turnover of the transport, relaxation and the linewidth as a function of disorder~\cite{Engelhardt2022,Chavez2021}. However,   due to the all-to-all coupling structure in single-mode models,  excitons can travel  instantaneously  between distant emitters  and thus exceed the  speed of light, potentially leading to an unphysical prediction for the transport efficiency. 

Since the photonic dispersion relation ensures the speed of light, the light fields should be described as a continuum of cavity  modes.   For example, the impact of disorder on polaritons was investigated perturbatively ~\cite{Izrailev1998,Agranovich2003,Litinskaya2006,Litinskaya2004}. Exact diagonalization and integration~\cite{Ribeiro2022,Allard2022}, mean-field based approaches~\cite{Patton2021,Cwik2014,Strashko2018,Sokolovskii2022}, Monte-Carlo methods~\cite{Xu2022} and density-functional theory~\cite{Alvertis2020}  have  been used to numerically investigate multimode models. Yet, a fully microscopic and analytical solution of the light-matter dynamics for disordered quantum emitters is still lacking.

\begin{figure*}[t]
	\includegraphics[width=\linewidth]{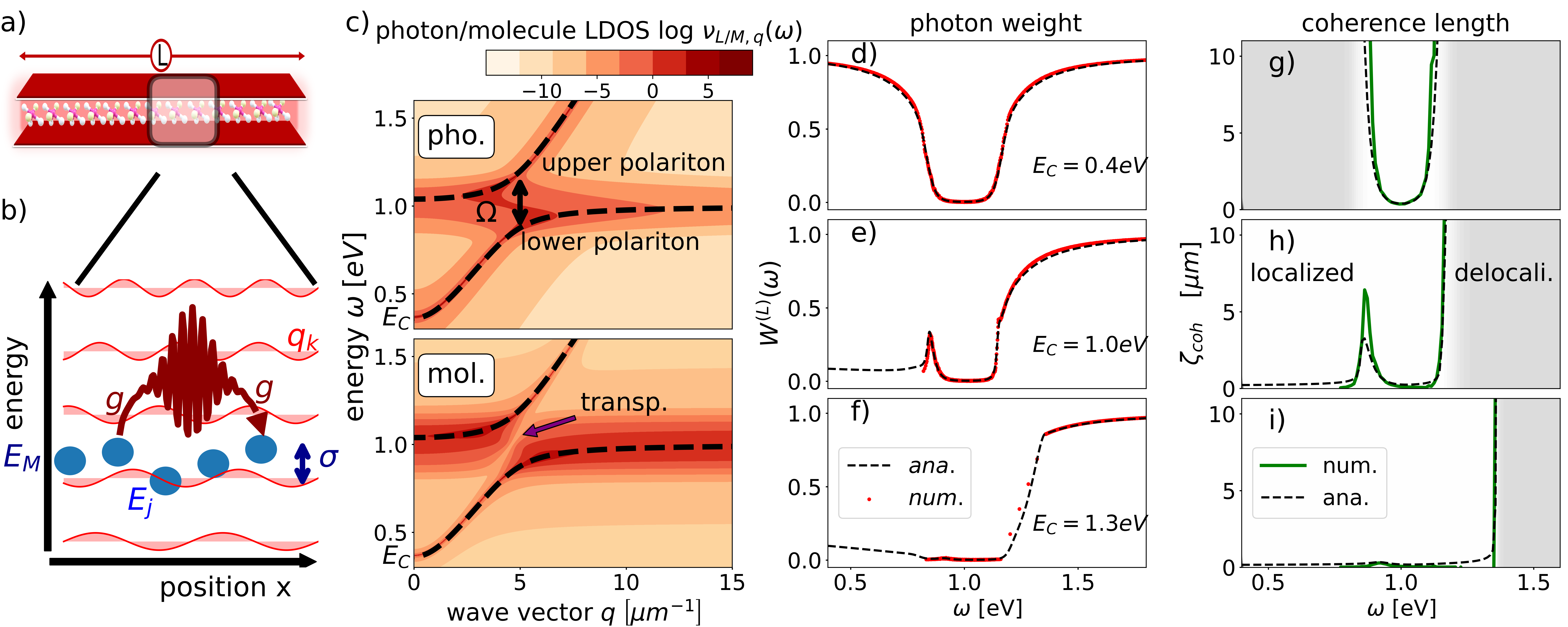}
	\caption{(a) One-dimensional microcavity of length $L$ containing $N$ molecules.  (b) Sketch of the energy configuration of the cavity modes (red sine functions)  and the molecules (blue circles, $E_j$ distributed around $E_{\text{M} }$ with Gaussian width $\sigma$). The molecules are coupled with strength $g_{j,k}$ to the photonic modes, such that  excitations can be transported  via photons. (c)  Wave-vector-resolved  photon and molecule LDOSs for $E_{C}=0.4\,\text{eV}$. (d),(e),(f) Average photon weight of the polaritons as a function of energy for $E_{C } = 0.4,1.0 ,1.3\,\text{eV}$. (g),(h)(i) Coherence length  of the polaritons for the same $E_{C }$ values as in (d),(e),(f).  Overall parameters are  $L = 125\,\mu \text{m}$,   $N= 5000$, $E_{\text{M} } = 1\,\text{eV}$, $\sigma = 0.05\,\text{eV}$, and $g\sqrt{\rho}= 0.14 \,\text{eV}$. The  photonic cutoff energy is $\omega_{\text{cut-off} }= 50 \,\text{eV}$, such  that  $5000$ photonic modes are included in the simulations. }
	\label{figOverview}
\end{figure*}

In this Letter, we  analytically and numerically solve   the multimode disordered Tavis-Cumming model   nonperturbatively.  Our closed-form solution predicts a finite coherence length for all polariton energies.
 Away from the average molecular energy $E_{\text{M} }$, the coherence length rapidly diverges and exceeds by far the typical length of realistic microcavities. This defines two transport regimes in the energy spectrum: one regime of strongly localized polaritons, where transport is diffusive, and one regime  of delocalized polaritons, where the large coherence length can support ballistic transport. The coherence length exhibits a turnover as a function of disorder, which has no analog in the Anderson localization~\cite{Anderson1958,Abrahams1979,Wang2020}, but is reminiscent of noise-assisted transport~\cite{Cao2009,Chuang2016}.

\textit{Multimode disordered Tavis-Cummings model.}---
 As shown in  Figs.~\ref{figOverview}(a) and~\ref{figOverview}(b),  we consider a one-dimensional  microcavity of length $L$ which contains $N$ quantum emitters representing  atoms, molecules, NV centers or particle-hole pairs in semiconductors. For concreteness, we focus on molecules in the following. We adopt  a multimode disordered Tavis-Cummings model, whose  Hamiltonian is given as $\hat H = \hat H_{\text{M}} + \hat H_{\text{L}} + \hat H_{\text{LM}},$
where 
\begin{eqnarray}
\hat H_{\text{M}} &=& \sum_{j=1}^{N} E_j \hat B_{j}^\dagger \hat B_j  \nonumber, \qquad
\hat H_{\text{L}} = \sum_{k} \;  \omega_{ k}   \hat a_{ k }^\dagger\hat a_{ k}  ,   \nonumber  \\   
\hat H_{\text{LM}} &=&   \sum_{j=1}^{N}\sum_{ k}      g_{j,  \kappa} \hat B_{j}^\dagger    \hat  a_{ k}  + \text{H.c.}  
\label{eq:hamiltonianTerms}
\end{eqnarray}
The molecules $j$  are described by bosonic  operators $\hat B_j $. Here, the excitation energies $E_j$ are distributed according to a Gaussian function
$
	P(E) =
		\frac{1}{\sqrt{\pi} \sigma} e^{-(E -E_{\text{M} })^2/(2\sigma^2) } , 
$
with center $E_{\text{M} }$ and disorder width $\sigma$. Yet, our findings also hold for arbitrary disorder distributions. The light field is quantized by the photonic operators $\hat a_{ k }$ labeled by $ k$. The photonic dispersion relation is $\omega_{k} =\sqrt{ c^2 q_{ k}^2 + E_{C}^2 } $, where $c$ is the speed of light, $q_k$ is the wave vector (specified below), and $E_{C} $ is  the  confinement energy depending on the  geometry of the microcavity. As the total excitation number $\hat n = \sum_j \hat B_j^\dagger\hat B_j  + \sum_k \hat a_{ k}^\dagger\hat a_{ k} $ is conserved, we can restrict our analysis to the single-excitation manifold. 
 The light-matter interaction in Eq.~\eqref{eq:hamiltonianTerms} is given by $g_{j,k} = g_{ k}   \varphi_{ k }( r_j )$ where $  r_j = j N/L$ is the position of  molecule $j$,  $g_{ k} $ is the wave vector dependent light-matter interaction, and     $\varphi_{ k }( r)$ are the photonic mode functions in one-dimensional space. We restrict the current investigation to energetic disorder, while spatial and orientational disorder will be considered elsewhere later.

  For the numerical calculations we use an open boundary condition, such that  the photonic modes are $\varphi_{ k } ( r) =  \sin\left( q_k r  \right)/\sqrt{L/2} $  for the  wave vectors $ q_{ k} = \pi k /L $ with integer $k>0$~\cite{supplementals}. In the analytical calculation, we assume a periodic boundary condition such that the photonic modes are  $\varphi_{ k } ( r) =  \exp\left(i q_k r  \right)/\sqrt{L} $, where $ q_{ k} =  2\pi k /L $ with integer $k$. We note that in the $L\rightarrow \infty$ limit, the boundary condition has a negligible effect.

\begin{figure}[t]
	\includegraphics[width=\linewidth]{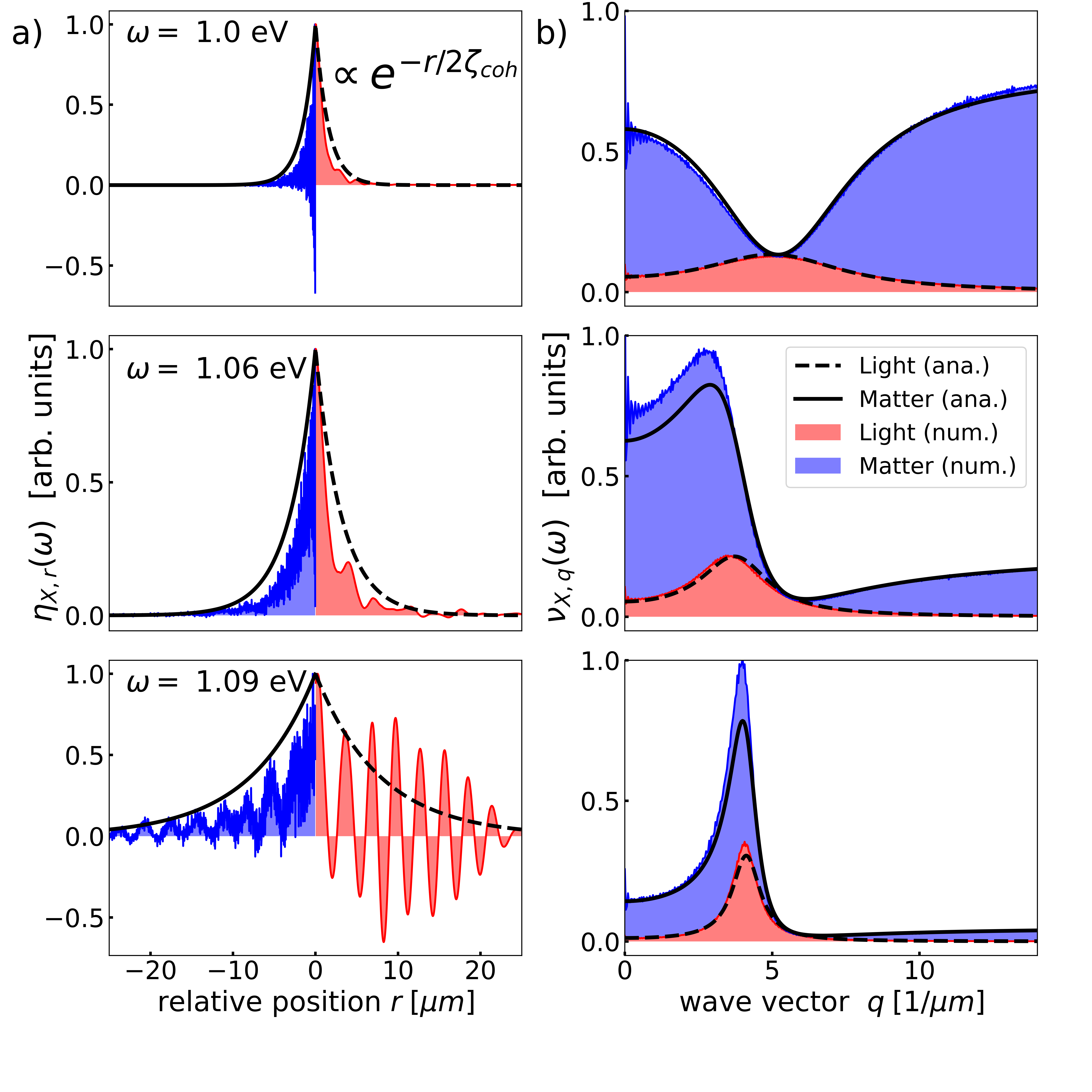}
	\caption{ (a) Imaginary part of the Green's function as a function of the relative position coordinate $r= r_{j} -r_{j'}$ as defined in Eq.~\eqref{eq:GreensFktImaginaryPart}. The results are shown in red (photon contribution, $r>0$) and blue (molecule contribution, $r<0$). The black lines depict the amplitude decay  predicted by the coherence length in Eq.~\eqref{eq:res:localiationLength}.  (b)  Imaginary part of the Green's function as a function of wave vector (i.e., the LDOS). The solid and dashed black lines depict the analytical predictions using Eq.~\eqref{eq:res:GreensFct}. Parameters are the same as in Fig.~\ref{figOverview} (d). Each Green's function has been averaged over an interval of width $\delta= 0.005 \left[\text{eV}\right]$. }
	\label{figWavefunctionAnalysis}
\end{figure}

\textit{Analytical solution.}
The Heisenberg equations of $\hat B_j  $ and $\hat a_{k} $ are transformed
into  the Laplace space defined by $\hat f(z) = \int_{0}^{\infty} dt e^{-zt} \hat f(t) $ for arbitrary   operators $\hat f(t)$. We find  that the coupling between different cavity modes $k _1,k_2$ scales as $\hat a_{k_1}(z) \propto \rho N^{-1/2} \hat a_{k _2}(z)$ and thus vanishes in the thermodynamic limit $N,L\rightarrow \infty$ with constant density $\rho=N/L$~\cite{supplementals}. In other words, one can treat the system as a  superpostion of uncoupled single-mode systems, which have been investigated in detail in Refs.~\cite{Engelhardt2022,Engelhardt2016a,Topp2015}. The solution of the Heisenberg operators in this limit is
\begin{eqnarray}
	\hat a_{k}(z)   &=& \frac{\hat a_{k}^{ (0)} }{ z + i \omega_{k} (z )  }  - i  \sum_j  \frac{g_{j,k } \hat B_j^{ (0)} }{\left[ z + i \omega_{k}(z )\right] \left(z + i E_j\right)}  \nonumber ,  \\
	 \hat B_j(z)   &=& \frac{\hat B_j^{ (0)}  }{z + i E_j} 
	   -i\sum_{k}  \frac{g_{j,k } \hat a_{k}^{ (0)} }{\left(z + i E_j\right) \left[ z + i \omega_{k}(z ) \right] } \nonumber  \\ 
	   &-&  \sum_{k}   \sum_{j_1}  \frac{ g_{j,k }  g_{j_1,k }^* \hat B_{j_1}^{ (0)}  }{\left(z + i E_j\right) \left[ z + i \omega_{k} (z ) \right] \left(z + i E_{j_1}\right)} , 
	\label{eq:operatorsLaplaceSpace}
\end{eqnarray}
where $\hat a_k^{ (0)} $ and $\hat B_j^{ (0)}  $ denote the initial conditions of the time evolution. We have defined the renormalized photon energy by
\begin{equation}
  \omega_{k} (z) =  \omega_{k }   - i \sum_{j }\frac{ \left|  g_{j,  k}\right|^2     }{z + i E_j} 
  \rightarrow   \omega_{k}    + g_{k}^2   \rho\Gamma(z)  ,
  \label{eq:frequencyNormalization}
\end{equation}
where  the $z$ dependence reflects a retardation effect. We have  expressed the disorder average in terms of the  density $\rho$ and the disorder-averaged Green's function of the unperturbed molecules  $\Gamma(z) =-i\int dE\left[  P(E)/ \left(z + i E \right) \right]  $. 
 Using  Eq.~\eqref{eq:operatorsLaplaceSpace}, we can construct arbitrary retarded Green's functions such as $G_{k,k^\prime }^{(\text{L})}(z) \equiv -i \left< \left[ \hat a_{k}(z),\hat a_{k^\prime}^{(0)\dagger}\right] \right>$ or  $G_{j,j'}^{(\text{M})}(z) \equiv- i \left< \left[ \hat B_j(z),\hat B_{j'}^{(0)\dagger}\right] \right>$.  Performing the disorder average, the  Green's function for $N,L\rightarrow \infty$ reads as
 \begin{eqnarray}
G_{k,k^\prime}^{(\text{L})}(z)  &=&  - i\frac{\delta_{k,k^\prime }}{z + i \omega_{k } (z ) }   \nonumber,  \\
 G_{j,j^\prime}^{(\text{M})}(z) &=&  \Gamma(z)\delta_{j,j^\prime}        -   i \sum_{k}\frac{g_{j,k }  g_{j_1,k }^*  }{ z + i\omega_{k } (z)}    \rho \Gamma(z )^2  .
 \label{eq:res:GreensFct}
 \end{eqnarray}
  These Green's functions  are equivalent to the single-mode system when the sum over $k$ is neglected. 
 The simple superposition of all $k$  modes reflects the mode decoupling in the thermodynamic limit, for which the matter system becomes homogeneous in a statistical sense.

\textit{Spectroscopy.}---
The wave-vector-resolved  photon and molecule local density of states (LDOSs) are given as  $\nu_{X,k}(\omega) \equiv-  \lim_{\delta\downarrow 0}\frac{1}{\pi}\text{Im}\, G_{k,k}^{(X)}(- i\omega +\delta) $ with $X=\text{L}$ and $X=\text{M}$, respectively, and can be measured spectroscopically~\cite{Engelhardt2022}.  

In Fig.~\ref{figOverview}(c), we investigate the  LDOSs for $E_{C} = 0.4\,\text{eV}$.  The LDOSs for $E_{C } = 1.0\,\text{eV}$ and $E_{C} =1.3\,\text{eV}$ can be found in the Supplementary Materials~\cite{supplementals}. The dashed lines depict the lower and upper polaritons  for a vanishing disorder $\sigma=0$. Close to  $\omega_{k} = E_{\text{M} }$, where both dispersions would cross for $g=0$, the lower and upper polaritons exhibit a Rabi splitting of $\Omega \approx 2 g \sqrt{ \rho } $. 
The photon and molecule LDOSs closely follow  the  photonic dispersion curves of the disorder-free systems (dashed). The photon LDOS accumulates close to the photon dispersion $\omega_{ k }$, but also around $E_{\text{M} }$ close to the polariton anticrossing, where light and matter are strongly mixed. The molecule LDOS accumulates around $E_{\text{M} }$, where it resembles the original disorder distribution. Along $\omega_{ k }$ and away from $E_{\text{M} }$, the molecule LDOS is one order of magnitude  smaller than the photon LDOS.
Because of  level repulsion, the molecule LDOS is suppressed for energies $\omega_{k }$ at the anticrossing (purple arrow), which resembles the electromagnetically induced transparency  and related effects~\cite{Fleischhauer2005,Engelhardt2022,Engelhardt2021,Herrera2022}. As each photon mode interacts with a disordered ensemble, the level repulsion is smeared out in the photon LDOS.

The photon  and molecule weights  of a specific eigenstate $\left|\alpha \right>$ with energy $\omega $ is given as $ W^{(X)}(\omega) \equiv \left< \alpha\right| \hat  P^{(X)} \left|\alpha \right> = \sum_{k} \nu_{X,k }(\omega) /\nu(\omega) $,  where $\nu(\omega) =\sum_{X=\text{L,M};k} \nu_{X,k}(\omega)  $, and $\hat  P^{(X)} $ is the photon (molecule) projection operator.
The  numerical calculation  in  Fig.~\ref{figOverview} (d,e,f)  verifies the analytical solution for various $E_{C}$.   (i) For $E_{C} = 0.4\,\text{eV}< E_{\text{M} } $, the photon weight vanishes around the resonance condition $ \omega \approx 1\,\text{eV}$, as the molecules by far outnumber the photon modes in this energy region. The photon weight increases monotonically with increasing distance  from the resonance condition.  (ii)  For  $E_{C } = 1.0\,\text{eV}=E_{\text{M} } $,  the photon weight does not monotonically increase with  distance from $E_{\text{M} }$. The peak around $\omega \approx 0.9\,\text{eV}$ is a consequence of the polariton formation, causing  the light field to be pushed down energetically. (iii) For $E_{C} = 1.3\,\text{eV} > E_{\text{M} }$, light and matter are energetically separated such that the mutual influence is rather weak. Motivated by Ref.~\cite{Ribeiro2022}, we define dark (bright) states as eigenstates with a photonic weight $W^{(\text{L})}<10\%$ ($W^{(\text{L})}>10\%$), which accumulate in the dip of the photon weight in Fig.~\ref{figOverview}(d).

\begin{figure}[t]
	\includegraphics[width=\linewidth]{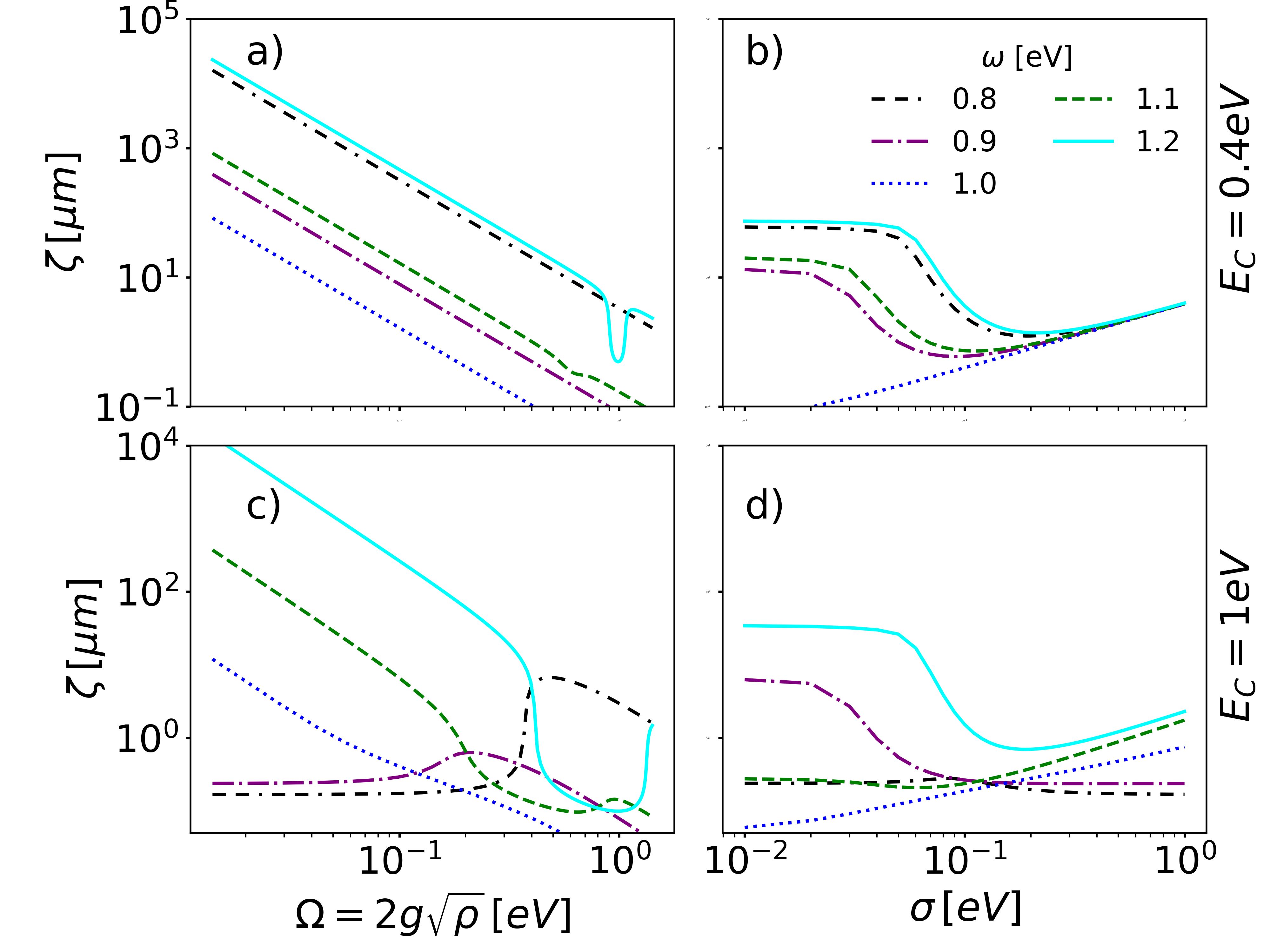}
	\caption{ Coherence length as a function of Rabi splitting [(a),(c)] and disorder [(b),(d)]. Parameters are the same as in Fig.~\ref{figOverview}. }
	\label{figAnalysisLocalizationLength}
\end{figure}

\textit{Polariton localization.}
Figure~\ref{figWavefunctionAnalysis} (a) depicts the imaginary part of the Green's function in position space,
\begin{eqnarray}
\eta_{X,r} (\omega)&\equiv&- \lim_{\delta\downarrow 0}\frac{1}{\pi}	\text{Im}\,  G_{j,j^\prime}^{(X)}  (-i\omega+\delta) \propto e^{-\frac{r}{2\zeta_{\text{coh} }}} ,
\label{eq:GreensFktImaginaryPart}
\end{eqnarray}
where $r= \left|r_j -r_{j^{\prime}} \right|$, for $E_{C} = 0.4 \;\text{eV}$ and three different energies $\omega$. In this definition we have used  the translational invariance of the Green's function  in the $N\rightarrow \infty$ limit. The photon (molecule) Green's function is depicted for $r>0$ ($r<0$).
Clearly, the amplitude of the Green's function  shows an exponential decay with increasing $r$, where the coherence length $\zeta_{\text{coh} }$ depends on energy. 

Figure~\ref{figWavefunctionAnalysis}(b) depicts the imaginary part of the Green's function in wave vector space, i.e., $\nu_{X,k}(\omega)$ for $X=\text{L,M}$. Overall, we observe that the widths of the Green's functions in position and wave vector space are related by the Heisenberg uncertainty principle. In contrast to the photon contribution, which converges to zero for large wave vectors $q$, the molecule Green's function converges to a finite value.  This is reflected by strong spatial fluctuations of the molecule Green's function in position space in Fig.~\ref{figWavefunctionAnalysis}(a), which are  absent in the photon Green's function. 

From Eq.~\eqref{eq:res:GreensFct} we can determine the coherence length  $\zeta_{\text{coh}}$ using  functional analysis, which characterizes the localization of the polaritons~\cite{Reed1975,supplementals}: In the $L\rightarrow \infty$ limit, we find
\begin{eqnarray}
\eta_{\text{L},r} (\omega)  \propto G_{j,j^\prime}^{(\text{L})}(-i\omega)&=&  \int dq\, G_q(\omega) e^{iqr}  ,
\label{eq:greensFktFourierTransformation}
\end{eqnarray}
where  $r= \left|r_j -r_{j^{\prime}} \right|\neq 0$ and $ G_q(\omega) = (-i)/\left[ -i\omega + i \omega_{qL/2\pi}( -i\omega)\right]$.
Specifically, the Green's function decays as $\propto e^{-\alpha r}$, where $\alpha $ is the largest value such that $G_{q-i\alpha'}$ is analytic for all $\left| \alpha' \right| <\alpha$.  $G_q$  has two types of non-analyticities, namely the roots of the denominator  and the branch cuts along the imaginary axis $\pm q \in \left[ i E_{C}/c, i \infty\right]$ due to the root in $\omega_{k}$. As explained later, the branch cut has minor influence on $\eta_{X,r} $, such that the coherence length is effectively determined by the root of $G_q$, i.e., 
\begin{eqnarray}
\zeta_{\text{coh} }^{-1} &   =  &  \frac 2{ c} \text{Im} \sqrt{ \left[ \omega -  g^2 \rho \Gamma(-i\omega) \right]^2 -E_{C }^2 },
\label{eq:res:localiationLength}
\end{eqnarray}
where  $g_{k} =g$ is assumed for simplicity.  Interestingly, the coherence length depends  via the product $g\sqrt{\rho}=\Omega/2$ (i.e., the Rabi frequency) on the light-matter coupling $g$.

In Figs.~\ref{figOverview}(g)-\ref{figOverview}(i) we compare the analytical expression for $\zeta_{\text{coh}}$ with the numerical evaluation~\cite{supplementals}, which  confirms the validity of the analytical solution. For large energies $\omega$, we observe that the coherence length diverges.  For realistic parameters and energies $\omega \approx E_{\text{M} }$, the branch cuts starting at $\pm i\frac{E_{C}}{c }$ have a minor influence on the coherence length, as for a large $E_{C}$,  $2c/E_{C}$ is significantly smaller than $\zeta_{\text{coh}} $, while for  small $E_{C}$, the influence of the branch cut in the Fourier transformation in Eq.~\eqref{eq:greensFktFourierTransformation} is negligible and the Green's function is still mainly determined by the pole of the Green's function~\cite{supplementals}.

\textit{Analysis.}---
In Fig.~\ref{figOverview} we demonstrate a correlation between the photon weight and the coherence length. As the  interaction between the molecules is mediated via photons, the coherence length increases when  photons can travel further without being scattered by molecules. The relation of  coherence length and photon scattering can be understood by expanding the Green's function in orders of $g$, where destructive interference of distinct photon scattering paths decrease the coherence length for increasing $g$~\cite{supplementals}. A low scattering probability  is reflected by a large photon weight in the Green's function. The coherence length of the Green's function can  thus be identified with the absorption length  for light traveling along the extended direction of the cavity according to Beer's absorption law~\cite{supplementals}.  Dark states have a detrimental impact on the coherence length. In general, we find a clear relation of dark states with a localized Green's function, and bright states with a delocalized Green's function. The localized (delocalized) regimes are thereby described by $\zeta_{\text{coh} }<\lambda_0$ ($\zeta_{\text{coh}}> \lambda_0$), where $\lambda_0 =hc/E_{\text{M} }$ is the resonance wavelength of the molecular excitations.

 In Fig.~\ref{figAnalysisLocalizationLength}, we analyze the coherence length $\zeta_{\text{coh}}$ as a function of $\Omega =2g\sqrt{\rho}$ and  $\sigma$ for $E_{C }= 0.4\,\text{eV}$ and $E_{C} = 1.0\,\text{eV}$.
 In Fig.~\ref{figAnalysisLocalizationLength}(a) for small $\Omega$, we observe  a clear linear dependence with slope $-2$ for all energies $\omega$. This can be explained by photon scattering, which consists of absorption ($\propto g\rho$) and reemission  ($\propto g$).
  Interestingly, the coherence length for $\omega=1.2\,\text{eV}$ exhibits a dip for large $\Omega$, as the matter LDOS is strongly deformed and accumulates around $\omega=1.2\,\text{eV}$, causing  enhanced photon scattering.  
 
 The observations  in Fig.~\ref{figAnalysisLocalizationLength}(c) for $E_{C} = 1.0\,\text{eV}$ and large $\omega= 1.0, 1.1, 1.2\,\text{eV}$ are qualitatively similar to  panel (a). The coherence length behaves very differently for small $\omega$, where the photonic modes are absent for $g=0$ as  $\omega_{k}> E_{C}$. As for these energies eigenstates can be only formed with  non-resonant photon modes,  the coherence length for small $\Omega$ is very small  and almost independent of $\Omega$~\cite{supplementals}. Interestingly, the coherence length  increases over more than 1 order of magnitude for $\Omega\approx 0.3$ and $\omega = 0.8\,\text{eV}$ because of the peak  in the photon  weight for small energies $\omega\approx0.8\,\text{eV}$  in Fig.~\ref{figOverview}(e).

  Analyzing the coherence length as a function of disorder in Fig.~\ref{figAnalysisLocalizationLength}(b), we observe a turnover as a function of $\sigma$. This  is in contrast to the  Anderson localization, where the coherence length monotonically decreases with disorder.   Recent work has revealed a turnover of the steady-state flux as a function of disorder  in the single-mode Tavis-Cummings  model~\cite{Engelhardt2022,Dubail2022,Chavez2021}, which can be explained by the overlap of the photon LDOS and the molecule energy disorder distribution $P(E)$~\cite{Engelhardt2022}. This interpretation can also be  employed here. For small $\sigma$, the  disorder distribution is strongly centered around $E_{\text{M} }$. With increasing $\sigma$, the disorder distribution increases for $\omega\neq E_{\text{M} }$, such that more molecules can resonantly scatter the photons with energy $\omega$, which reduces the coherence length. For a large disorder, the molecule energies spread over a large energy regime, such that there are only few molecules  in resonance with the photon modes close to $\omega$, which enhances the coherence length. As the Gauss distribution becomes very flat close to the center  for large $\sigma$, the coherence length becomes independent of  $\omega$ for large $\sigma$. For  $\omega = 1.0\,\text{eV}$, we do not observe a turnover, as the disorder distribution $P(\omega \approx E_{\text{M} })$ decreases monotonically for increasing $\sigma$. 
The turnovers can be also observed  for $E_{C}= 1.0\,\text{eV}$ in Fig.~\ref{figAnalysisLocalizationLength}(d) for large $\omega=  1.1, 1.2\,\text{eV}$, while overall the dependence on $\sigma$ is more complicated because of the significant influence of the square root  dispersion relation of $\omega_{k  }$ close to $q=0$.

\textit{Conclusions.}---
 We have analytically and numerically solved the multimode disordered Tavis-Cummings model and predict its dispersion and localization properties.  (i) The analytical solution is built on the mode decoupling and statistical self-averaging and is exact in the thermodynamic limit. Based on the solution, wave-vector-resolved properties such as broadened spectral line shape and dispersion can be predicted effectively within the single-mode treatment, whereas spatial-dependent properties such as transport and coherence length involve a wave vector summation and thus require the multimode formalism.  (ii)  A coherence length  is introduced to characterize the finite size of the eigenstates as a function of the excitation energy and shows transitions from localized states around the molecular energy ($E_{\text{M} }$) to delocalized states away from $E_{\text{M} }$. These transitions are strongly correlated with the photon weight and define a ballistic and a localized transport regime. (iii) Intriguingly, the coherence length is inversely proportional to the square of the Rabi frequency and can exhibit a turnover as a function of disorder. (iv) Both the dispersion and coherence length depend strongly on the cavity confinement energy $E_{C}$: the number of available resonant photon modes and thus the light-matter coupling regime increase as the cavity changes from blueshifted ($E_{C}>E_{\text{M} }$),  resonance ($E_{C }=E_{\text{M} }$), to redshifted ($E_{C}<E_{\text{M} }$).  The current investigation focuses on the one-dimensional system with energetic disorder, while higher-dimensional systems with spatial and orientational disorder will be considered elsewhere.

The coherence length crucially depends  on the  light-matter coupling and the disorder. For example,  it can be enhanced by more than one order of magnitude with a slight increase of the light-matter interaction [cf. Fig.~\ref{figAnalysisLocalizationLength}(c)]. Moreover, it can exhibit a turnover as a function of disorder, which contrasts the monotonically decreasing coherence length known from the  Anderson localization, but is reminiscent of noise-assisted quantum transport~\cite{Wu2013,Lee2015,Engelhardt2019a,Chenu2017}. Arising from the overlap of the light LDOS and the  disorder distribution, this turnover is induced by the same mechanism as the transport turnover previously predicted in the single-mode disordered Tavis-Cummings model~\cite{Engelhardt2022}. Experimentally, this turnover can be investigated using a mixture of two molecular ensembles as in \cite{Cohn2022}.

Noteworthy, the experiment in Ref.~\cite{Balasubrahmaniyam2023} has identified a transition from  diffusive transport for small photonic weight  to ballistic transport for large photonic weight. 
This observation is in perfect agreement with our analytical calculation, which predicts localized (i.e., diffusive) and delocalized (i.e., ballistic) eigenstates and a sharp transition as a function of excitation energy, as shown in
 Fig.~\ref{figOverview}. Theses findings reveal that the photonic weight explains the enhanced transport efficiency. In general, dark states with low photon weight correspond to localized states, while bright states with high photon weight correspond to delocalized states. 

The experiment in Ref.~\cite{Xu2022} indicates that phonon-assisted coupling of diffusive eigenstates and ballistic eigenstates helps to overcome the localization. Extending our current model in Eq.~\eqref{eq:hamiltonianTerms} to incorporate phonon modes will quantitatively demonstrate this mechanism.
Moreover, as experimentally shown  in~\cite{Pandya2022}, the detrimental impact of the cavity quality on transport properties can be modeled  by a complex dispersion relation $\omega_{k }\rightarrow \omega_{k } -i\kappa $ with $\kappa>0$. This will result in a complex energy shift $\omega\rightarrow \omega+i\kappa $ in the coherence length in Eq.~\eqref{eq:res:localiationLength}, leading to a suppression of transport. These and other experimental implications will be studied in future works. \\

 G. E. gratefully acknowledges financial support from the Guangdong Provincial Key Laboratory of Quantum Science and Engineering (Grant No.2019B121203002); J. C. acknowledges support from the NSF (Grants No. CHE 1800301 and No. CHE1836913), the MIT sloan fund, and the Maria Curie FRIAS COFUND Fellowship Programme (FCFP) during his sabbatical in Germany. 

\bibliography{mybibliography}

\begin{thebibliography}{56}%
\makeatletter
\providecommand \@ifxundefined [1]{%
 \@ifx{#1\undefined}
}%
\providecommand \@ifnum [1]{%
 \ifnum #1\expandafter \@firstoftwo
 \else \expandafter \@secondoftwo
 \fi
}%
\providecommand \@ifx [1]{%
 \ifx #1\expandafter \@firstoftwo
 \else \expandafter \@secondoftwo
 \fi
}%
\providecommand \natexlab [1]{#1}%
\providecommand \enquote  [1]{``#1''}%
\providecommand \bibnamefont  [1]{#1}%
\providecommand \bibfnamefont [1]{#1}%
\providecommand \citenamefont [1]{#1}%
\providecommand \href@noop [0]{\@secondoftwo}%
\providecommand \href [0]{\begingroup \@sanitize@url \@href}%
\providecommand \@href[1]{\@@startlink{#1}\@@href}%
\providecommand \@@href[1]{\endgroup#1\@@endlink}%
\providecommand \@sanitize@url [0]{\catcode `\\12\catcode `\$12\catcode
  `\&12\catcode `\#12\catcode `\^12\catcode `\_12\catcode `\%12\relax}%
\providecommand \@@startlink[1]{}%
\providecommand \@@endlink[0]{}%
\providecommand \url  [0]{\begingroup\@sanitize@url \@url }%
\providecommand \@url [1]{\endgroup\@href {#1}{\urlprefix }}%
\providecommand \urlprefix  [0]{URL }%
\providecommand \Eprint [0]{\href }%
\providecommand \doibase [0]{https://doi.org/}%
\providecommand \selectlanguage [0]{\@gobble}%
\providecommand \bibinfo  [0]{\@secondoftwo}%
\providecommand \bibfield  [0]{\@secondoftwo}%
\providecommand \translation [1]{[#1]}%
\providecommand \BibitemOpen [0]{}%
\providecommand \bibitemStop [0]{}%
\providecommand \bibitemNoStop [0]{.\EOS\space}%
\providecommand \EOS [0]{\spacefactor3000\relax}%
\providecommand \BibitemShut  [1]{\csname bibitem#1\endcsname}%
\let\auto@bib@innerbib\@empty
\bibitem [{\citenamefont {Garcia-Vidal}\ \emph {et~al.}(2021)\citenamefont
  {Garcia-Vidal}, \citenamefont {Ciuti},\ and\ \citenamefont
  {Ebbesen}}]{Garcia-Vidal2021}%
  \BibitemOpen
  \bibfield  {author} {\bibinfo {author} {\bibfnamefont {F.~J.}\ \bibnamefont
  {Garcia-Vidal}}, \bibinfo {author} {\bibfnamefont {C.}~\bibnamefont
  {Ciuti}},\ and\ \bibinfo {author} {\bibfnamefont {T.~W.}\ \bibnamefont
  {Ebbesen}},\ }\bibfield  {title} {\bibinfo {title} {Manipulating matter by
  strong coupling to vacuum fields},\ }\href
  {https://doi.org/10.1126/science.abd0336} {\bibfield  {journal} {\bibinfo
  {journal} {Science}\ }\textbf {\bibinfo {volume} {373}},\ \bibinfo {pages}
  {0336} (\bibinfo {year} {2021})}\BibitemShut {NoStop}%
\bibitem [{\citenamefont {Lerario}\ \emph {et~al.}(2017)\citenamefont
  {Lerario}, \citenamefont {Ballarini}, \citenamefont {Fieramosca},
  \citenamefont {Cannavale}, \citenamefont {Genco}, \citenamefont {Mangione},
  \citenamefont {Gambino}, \citenamefont {Dominici}, \citenamefont {De~Giorgi},
  \citenamefont {Gigli},\ and\ \citenamefont {Sanvitto}}]{Lerario2017}%
  \BibitemOpen
  \bibfield  {author} {\bibinfo {author} {\bibfnamefont {G.}~\bibnamefont
  {Lerario}}, \bibinfo {author} {\bibfnamefont {D.}~\bibnamefont {Ballarini}},
  \bibinfo {author} {\bibfnamefont {A.}~\bibnamefont {Fieramosca}}, \bibinfo
  {author} {\bibfnamefont {A.}~\bibnamefont {Cannavale}}, \bibinfo {author}
  {\bibfnamefont {A.}~\bibnamefont {Genco}}, \bibinfo {author} {\bibfnamefont
  {F.}~\bibnamefont {Mangione}}, \bibinfo {author} {\bibfnamefont
  {S.}~\bibnamefont {Gambino}}, \bibinfo {author} {\bibfnamefont
  {L.}~\bibnamefont {Dominici}}, \bibinfo {author} {\bibfnamefont
  {M.}~\bibnamefont {De~Giorgi}}, \bibinfo {author} {\bibfnamefont
  {G.}~\bibnamefont {Gigli}},\ and\ \bibinfo {author} {\bibfnamefont
  {D.}~\bibnamefont {Sanvitto}},\ }\bibfield  {title} {\bibinfo {title}
  {High-speed flow of interacting organic polaritons},\ }\href
  {https://doi.org/10.1038/lsa.2016.212} {\bibfield  {journal} {\bibinfo
  {journal} {Light: Sci. Appl.}\ }\textbf {\bibinfo {volume} {6}},\ \bibinfo
  {pages} {e16212} (\bibinfo {year} {2017})}\BibitemShut {NoStop}%
\bibitem [{\citenamefont {Orgiu}\ \emph {et~al.}(2015)\citenamefont {Orgiu},
  \citenamefont {George}, \citenamefont {Hutchison}, \citenamefont {Devaux},
  \citenamefont {Dayen}, \citenamefont {Doudin}, \citenamefont {Stellacci},
  \citenamefont {Genet}, \citenamefont {Schachenmayer}, \citenamefont {Genes},
  \citenamefont {Pupillo}, \citenamefont {Samorì},\ and\ \citenamefont
  {Ebbesen}}]{Orgiu2015}%
  \BibitemOpen
  \bibfield  {author} {\bibinfo {author} {\bibfnamefont {E.}~\bibnamefont
  {Orgiu}}, \bibinfo {author} {\bibfnamefont {J.}~\bibnamefont {George}},
  \bibinfo {author} {\bibfnamefont {J.~A.}\ \bibnamefont {Hutchison}}, \bibinfo
  {author} {\bibfnamefont {E.}~\bibnamefont {Devaux}}, \bibinfo {author}
  {\bibfnamefont {J.~F.}\ \bibnamefont {Dayen}}, \bibinfo {author}
  {\bibfnamefont {B.}~\bibnamefont {Doudin}}, \bibinfo {author} {\bibfnamefont
  {F.}~\bibnamefont {Stellacci}}, \bibinfo {author} {\bibfnamefont
  {C.}~\bibnamefont {Genet}}, \bibinfo {author} {\bibfnamefont
  {J.}~\bibnamefont {Schachenmayer}}, \bibinfo {author} {\bibfnamefont
  {C.}~\bibnamefont {Genes}}, \bibinfo {author} {\bibfnamefont
  {G.}~\bibnamefont {Pupillo}}, \bibinfo {author} {\bibfnamefont
  {P.}~\bibnamefont {Samorì}},\ and\ \bibinfo {author} {\bibfnamefont {T.~W.}\
  \bibnamefont {Ebbesen}},\ }\bibfield  {title} {\bibinfo {title} {Conductivity
  in organic semiconductors hybridized with the vacuum field},\ }\href
  {https://doi.org/10.1038/nmat4392} {\bibfield  {journal} {\bibinfo  {journal}
  {Nat. Mater.}\ }\textbf {\bibinfo {volume} {14}},\ \bibinfo {pages} {1123}
  (\bibinfo {year} {2015})}\BibitemShut {NoStop}%
\bibitem [{\citenamefont {Rozenman}\ \emph {et~al.}(2018)\citenamefont
  {Rozenman}, \citenamefont {Akulov}, \citenamefont {Golombek},\ and\
  \citenamefont {Schwartz}}]{Rozenman2018}%
  \BibitemOpen
  \bibfield  {author} {\bibinfo {author} {\bibfnamefont {G.~G.}\ \bibnamefont
  {Rozenman}}, \bibinfo {author} {\bibfnamefont {K.}~\bibnamefont {Akulov}},
  \bibinfo {author} {\bibfnamefont {A.}~\bibnamefont {Golombek}},\ and\
  \bibinfo {author} {\bibfnamefont {T.}~\bibnamefont {Schwartz}},\ }\bibfield
  {title} {\bibinfo {title} {Long-range transport of organic exciton-polaritons
  revealed by ultrafast microscopy},\ }\href
  {https://doi.org/10.1021/acsphotonics.7b01332} {\bibfield  {journal}
  {\bibinfo  {journal} {ACS Photonics}\ }\textbf {\bibinfo {volume} {5}},\
  \bibinfo {pages} {105} (\bibinfo {year} {2018})}\BibitemShut {NoStop}%
\bibitem [{\citenamefont {Hou}\ \emph {et~al.}(2020)\citenamefont {Hou},
  \citenamefont {Khatoniar}, \citenamefont {Ding}, \citenamefont {Qu},
  \citenamefont {Napolov}, \citenamefont {Menon},\ and\ \citenamefont
  {Forrest}}]{Hou2020}%
  \BibitemOpen
  \bibfield  {author} {\bibinfo {author} {\bibfnamefont {S.}~\bibnamefont
  {Hou}}, \bibinfo {author} {\bibfnamefont {M.}~\bibnamefont {Khatoniar}},
  \bibinfo {author} {\bibfnamefont {K.}~\bibnamefont {Ding}}, \bibinfo {author}
  {\bibfnamefont {Y.}~\bibnamefont {Qu}}, \bibinfo {author} {\bibfnamefont
  {A.}~\bibnamefont {Napolov}}, \bibinfo {author} {\bibfnamefont {V.~M.}\
  \bibnamefont {Menon}},\ and\ \bibinfo {author} {\bibfnamefont {S.~R.}\
  \bibnamefont {Forrest}},\ }\bibfield  {title} {\bibinfo {title}
  {Ultralong-range energy transport in a disordered organic semiconductor at
  room temperature via coherent exciton-polariton propagation},\ }\href@noop {}
  {\bibfield  {journal} {\bibinfo  {journal} {Adv. Mater.}\ }\textbf {\bibinfo
  {volume} {32}},\ \bibinfo {pages} {2002127} (\bibinfo {year}
  {2020})}\BibitemShut {NoStop}%
\bibitem [{\citenamefont {Krainova}\ \emph {et~al.}(2020)\citenamefont
  {Krainova}, \citenamefont {Grede}, \citenamefont {Tsokkou}, \citenamefont
  {Banerji},\ and\ \citenamefont {Giebink}}]{Krainova2020}%
  \BibitemOpen
  \bibfield  {author} {\bibinfo {author} {\bibfnamefont {N.}~\bibnamefont
  {Krainova}}, \bibinfo {author} {\bibfnamefont {A.~J.}\ \bibnamefont {Grede}},
  \bibinfo {author} {\bibfnamefont {D.}~\bibnamefont {Tsokkou}}, \bibinfo
  {author} {\bibfnamefont {N.}~\bibnamefont {Banerji}},\ and\ \bibinfo {author}
  {\bibfnamefont {N.~C.}\ \bibnamefont {Giebink}},\ }\bibfield  {title}
  {\bibinfo {title} {{Polaron Photoconductivity in the Weak and Strong
  Light-Matter Coupling Regime}},\ }\href
  {https://doi.org/10.1103/PhysRevLett.124.177401} {\bibfield  {journal}
  {\bibinfo  {journal} {Phys. Rev. Lett.}\ }\textbf {\bibinfo {volume} {124}},\
  \bibinfo {pages} {177401} (\bibinfo {year} {2020})}\BibitemShut {NoStop}%
\bibitem [{\citenamefont {Balasubrahmaniyam}\ \emph {et~al.}(2023)\citenamefont
  {Balasubrahmaniyam}, \citenamefont {Simkhovich}, \citenamefont {Golombek},
  \citenamefont {Sandik}, \citenamefont {Ankonina},\ and\ \citenamefont
  {Schwartz}}]{Balasubrahmaniyam2023}%
  \BibitemOpen
  \bibfield  {author} {\bibinfo {author} {\bibfnamefont {M.}~\bibnamefont
  {Balasubrahmaniyam}}, \bibinfo {author} {\bibfnamefont {A.}~\bibnamefont
  {Simkhovich}}, \bibinfo {author} {\bibfnamefont {A.}~\bibnamefont
  {Golombek}}, \bibinfo {author} {\bibfnamefont {G.}~\bibnamefont {Sandik}},
  \bibinfo {author} {\bibfnamefont {G.}~\bibnamefont {Ankonina}},\ and\
  \bibinfo {author} {\bibfnamefont {T.}~\bibnamefont {Schwartz}},\ }\bibfield
  {title} {\bibinfo {title} {From enhanced diffusion to ultrafast ballistic
  motion of hybrid light-matter excitations},\ }\href
  {https://doi.org/10.1038/s41563-022-01463-3} {\bibfield  {journal} {\bibinfo
  {journal} {Nat. Mater.}\ }\textbf {\bibinfo {volume} {22}},\ \bibinfo {pages}
  {338} (\bibinfo {year} {2023})}\BibitemShut {NoStop}%
\bibitem [{\citenamefont {Ch\'avez}\ \emph {et~al.}(2021)\citenamefont
  {Ch\'avez}, \citenamefont {Mattiotti}, \citenamefont {M\'endez-Berm\'udez},
  \citenamefont {Borgonovi},\ and\ \citenamefont {Celardo}}]{Chavez2021}%
  \BibitemOpen
  \bibfield  {author} {\bibinfo {author} {\bibfnamefont {N.~C.}\ \bibnamefont
  {Ch\'avez}}, \bibinfo {author} {\bibfnamefont {F.}~\bibnamefont {Mattiotti}},
  \bibinfo {author} {\bibfnamefont {J.~A.}\ \bibnamefont
  {M\'endez-Berm\'udez}}, \bibinfo {author} {\bibfnamefont {F.}~\bibnamefont
  {Borgonovi}},\ and\ \bibinfo {author} {\bibfnamefont {G.~L.}\ \bibnamefont
  {Celardo}},\ }\bibfield  {title} {\bibinfo {title} {{Disorder-Enhanced and
  Disorder-Independent Transport with Long-Range Hopping: Application to
  Molecular Chains in Optical Cavities}},\ }\href
  {https://doi.org/10.1103/PhysRevLett.126.153201} {\bibfield  {journal}
  {\bibinfo  {journal} {Phys. Rev. Lett.}\ }\textbf {\bibinfo {volume} {126}},\
  \bibinfo {pages} {153201} (\bibinfo {year} {2021})}\BibitemShut {NoStop}%
\bibitem [{\citenamefont {Engelhardt}\ and\ \citenamefont
  {Cao}(2022)}]{Engelhardt2022}%
  \BibitemOpen
  \bibfield  {author} {\bibinfo {author} {\bibfnamefont {G.}~\bibnamefont
  {Engelhardt}}\ and\ \bibinfo {author} {\bibfnamefont {J.}~\bibnamefont
  {Cao}},\ }\bibfield  {title} {\bibinfo {title} {Unusual dynamical properties
  of disordered polaritons in microcavities},\ }\href
  {https://doi.org/10.1103/PhysRevB.105.064205} {\bibfield  {journal} {\bibinfo
   {journal} {Phys. Rev. B}\ }\textbf {\bibinfo {volume} {105}},\ \bibinfo
  {pages} {064205} (\bibinfo {year} {2022})}\BibitemShut {NoStop}%
\bibitem [{\citenamefont {Feist}\ and\ \citenamefont
  {Garcia-Vidal}(2015)}]{Feist2015}%
  \BibitemOpen
  \bibfield  {author} {\bibinfo {author} {\bibfnamefont {J.}~\bibnamefont
  {Feist}}\ and\ \bibinfo {author} {\bibfnamefont {F.~J.}\ \bibnamefont
  {Garcia-Vidal}},\ }\bibfield  {title} {\bibinfo {title} {{Extraordinary
  Exciton Conductance Induced by Strong Coupling}},\ }\href
  {https://doi.org/10.1103/PhysRevLett.114.196402} {\bibfield  {journal}
  {\bibinfo  {journal} {Phys. Rev. Lett.}\ }\textbf {\bibinfo {volume} {114}},\
  \bibinfo {pages} {196402} (\bibinfo {year} {2015})}\BibitemShut {NoStop}%
\bibitem [{\citenamefont {Sommer}\ \emph {et~al.}(2021)\citenamefont {Sommer},
  \citenamefont {Reitz}, \citenamefont {Mineo},\ and\ \citenamefont
  {Genes}}]{Sommer2021}%
  \BibitemOpen
  \bibfield  {author} {\bibinfo {author} {\bibfnamefont {C.}~\bibnamefont
  {Sommer}}, \bibinfo {author} {\bibfnamefont {M.}~\bibnamefont {Reitz}},
  \bibinfo {author} {\bibfnamefont {F.}~\bibnamefont {Mineo}},\ and\ \bibinfo
  {author} {\bibfnamefont {C.}~\bibnamefont {Genes}},\ }\bibfield  {title}
  {\bibinfo {title} {Molecular polaritonics in dense mesoscopic disordered
  ensembles},\ }\href {https://doi.org/10.1103/PhysRevResearch.3.033141}
  {\bibfield  {journal} {\bibinfo  {journal} {Phys. Rev. Res.}\ }\textbf
  {\bibinfo {volume} {3}},\ \bibinfo {pages} {033141} (\bibinfo {year}
  {2021})}\BibitemShut {NoStop}%
\bibitem [{\citenamefont {Spano}(2015)}]{Spano2015}%
  \BibitemOpen
  \bibfield  {author} {\bibinfo {author} {\bibfnamefont {F.~C.}\ \bibnamefont
  {Spano}},\ }\bibfield  {title} {\bibinfo {title} {Optical microcavities
  enhance the exciton coherence length and eliminate vibronic coupling in
  {J}-aggregates},\ }\href {https://doi.org/10.1063/1.4919348} {\bibfield
  {journal} {\bibinfo  {journal} {J. Chem. Phys.}\ }\textbf {\bibinfo {volume}
  {142}},\ \bibinfo {pages} {184707} (\bibinfo {year} {2015})}\BibitemShut
  {NoStop}%
\bibitem [{\citenamefont {Shammah}\ \emph {et~al.}(2017)\citenamefont
  {Shammah}, \citenamefont {Lambert}, \citenamefont {Nori},\ and\ \citenamefont
  {De~Liberato}}]{Shammah2017}%
  \BibitemOpen
  \bibfield  {author} {\bibinfo {author} {\bibfnamefont {N.}~\bibnamefont
  {Shammah}}, \bibinfo {author} {\bibfnamefont {N.}~\bibnamefont {Lambert}},
  \bibinfo {author} {\bibfnamefont {F.}~\bibnamefont {Nori}},\ and\ \bibinfo
  {author} {\bibfnamefont {S.}~\bibnamefont {De~Liberato}},\ }\bibfield
  {title} {\bibinfo {title} {Superradiance with local phase-breaking effects},\
  }\href {https://doi.org/10.1103/PhysRevA.96.023863} {\bibfield  {journal}
  {\bibinfo  {journal} {Phys. Rev. A}\ }\textbf {\bibinfo {volume} {96}},\
  \bibinfo {pages} {023863} (\bibinfo {year} {2017})}\BibitemShut {NoStop}%
\bibitem [{\citenamefont {Botzung}\ \emph {et~al.}(2020)\citenamefont
  {Botzung}, \citenamefont {Hagenm\"uller}, \citenamefont {Sch\"utz},
  \citenamefont {Dubail}, \citenamefont {Pupillo},\ and\ \citenamefont
  {Schachenmayer}}]{Botzung2020}%
  \BibitemOpen
  \bibfield  {author} {\bibinfo {author} {\bibfnamefont {T.}~\bibnamefont
  {Botzung}}, \bibinfo {author} {\bibfnamefont {D.}~\bibnamefont
  {Hagenm\"uller}}, \bibinfo {author} {\bibfnamefont {S.}~\bibnamefont
  {Sch\"utz}}, \bibinfo {author} {\bibfnamefont {J.}~\bibnamefont {Dubail}},
  \bibinfo {author} {\bibfnamefont {G.}~\bibnamefont {Pupillo}},\ and\ \bibinfo
  {author} {\bibfnamefont {J.}~\bibnamefont {Schachenmayer}},\ }\bibfield
  {title} {\bibinfo {title} {Dark state semilocalization of quantum emitters in
  a cavity},\ }\href {https://doi.org/10.1103/PhysRevB.102.144202} {\bibfield
  {journal} {\bibinfo  {journal} {Phys. Rev. B}\ }\textbf {\bibinfo {volume}
  {102}},\ \bibinfo {pages} {144202} (\bibinfo {year} {2020})}\BibitemShut
  {NoStop}%
\bibitem [{\citenamefont {Dubail}\ \emph {et~al.}(2022)\citenamefont {Dubail},
  \citenamefont {Botzung}, \citenamefont {Schachenmayer}, \citenamefont
  {Pupillo},\ and\ \citenamefont {Hagenm\"uller}}]{Dubail2022}%
  \BibitemOpen
  \bibfield  {author} {\bibinfo {author} {\bibfnamefont {J.}~\bibnamefont
  {Dubail}}, \bibinfo {author} {\bibfnamefont {T.}~\bibnamefont {Botzung}},
  \bibinfo {author} {\bibfnamefont {J.}~\bibnamefont {Schachenmayer}}, \bibinfo
  {author} {\bibfnamefont {G.}~\bibnamefont {Pupillo}},\ and\ \bibinfo {author}
  {\bibfnamefont {D.}~\bibnamefont {Hagenm\"uller}},\ }\bibfield  {title}
  {\bibinfo {title} {Large random arrowhead matrices: Multifractality,
  semilocalization, and protected transport in disordered quantum spins coupled
  to a cavity},\ }\href {https://doi.org/10.1103/PhysRevA.105.023714}
  {\bibfield  {journal} {\bibinfo  {journal} {Phys. Rev. A}\ }\textbf {\bibinfo
  {volume} {105}},\ \bibinfo {pages} {023714} (\bibinfo {year}
  {2022})}\BibitemShut {NoStop}%
\bibitem [{\citenamefont {Zhang}\ and\ \citenamefont
  {Zhang}(2021)}]{Zhang2021}%
  \BibitemOpen
  \bibfield  {author} {\bibinfo {author} {\bibfnamefont {Q.}~\bibnamefont
  {Zhang}}\ and\ \bibinfo {author} {\bibfnamefont {K.}~\bibnamefont {Zhang}},\
  }\bibfield  {title} {\bibinfo {title} {{Collective effects of organic
  molecules based on the Holstein–Tavis–Cummings model}},\ }\href
  {https://doi.org/10.1088/1361-6455/ac0afa} {\bibfield  {journal} {\bibinfo
  {journal} {J. Phys. B}\ }\textbf {\bibinfo {volume} {54}},\ \bibinfo {pages}
  {145101} (\bibinfo {year} {2021})}\BibitemShut {NoStop}%
\bibitem [{\citenamefont {Gera}\ and\ \citenamefont
  {Sebastian}(2022)}]{Gera2022a}%
  \BibitemOpen
  \bibfield  {author} {\bibinfo {author} {\bibfnamefont {T.}~\bibnamefont
  {Gera}}\ and\ \bibinfo {author} {\bibfnamefont {K.~L.}\ \bibnamefont
  {Sebastian}},\ }\bibfield  {title} {\bibinfo {title} {Exact results for the
  {Tavis-Cummings} and {Hückel Hamiltonians} with diagonal disorder},\ }\href
  {https://doi.org/10.1021/acs.jpca.2c02359} {\bibfield  {journal} {\bibinfo
  {journal} {J. Phys. Chem. A}\ }\textbf {\bibinfo {volume} {126}},\ \bibinfo
  {pages} {5449} (\bibinfo {year} {2022})}\BibitemShut {NoStop}%
\bibitem [{\citenamefont {Cohn}\ \emph {et~al.}(2022)\citenamefont {Cohn},
  \citenamefont {Sufrin}, \citenamefont {Basu},\ and\ \citenamefont
  {Chuntonov}}]{Cohn2022}%
  \BibitemOpen
  \bibfield  {author} {\bibinfo {author} {\bibfnamefont {B.}~\bibnamefont
  {Cohn}}, \bibinfo {author} {\bibfnamefont {S.}~\bibnamefont {Sufrin}},
  \bibinfo {author} {\bibfnamefont {A.}~\bibnamefont {Basu}},\ and\ \bibinfo
  {author} {\bibfnamefont {L.}~\bibnamefont {Chuntonov}},\ }\bibfield  {title}
  {\bibinfo {title} {Vibrational polaritons in disordered molecular
  ensembles},\ }\href {https://doi.org/10.1021/acs.jpclett.2c02341} {\bibfield
  {journal} {\bibinfo  {journal} {J. Phys. Chem. Lett.}\ }\textbf {\bibinfo
  {volume} {13}},\ \bibinfo {pages} {8369} (\bibinfo {year}
  {2022})}\BibitemShut {NoStop}%
\bibitem [{\citenamefont {Herrera}\ and\ \citenamefont
  {Spano}(2016)}]{Herrera2016}%
  \BibitemOpen
  \bibfield  {author} {\bibinfo {author} {\bibfnamefont {F.}~\bibnamefont
  {Herrera}}\ and\ \bibinfo {author} {\bibfnamefont {F.~C.}\ \bibnamefont
  {Spano}},\ }\bibfield  {title} {\bibinfo {title} {{Cavity-Controlled
  Chemistry in Molecular Ensembles}},\ }\href
  {https://doi.org/10.1103/PhysRevLett.116.238301} {\bibfield  {journal}
  {\bibinfo  {journal} {Phys. Rev. Lett.}\ }\textbf {\bibinfo {volume} {116}},\
  \bibinfo {pages} {238301} (\bibinfo {year} {2016})}\BibitemShut {NoStop}%
\bibitem [{\citenamefont {Houdr\'e}\ \emph {et~al.}(1996)\citenamefont
  {Houdr\'e}, \citenamefont {Stanley},\ and\ \citenamefont
  {Ilegems}}]{Houdre1996}%
  \BibitemOpen
  \bibfield  {author} {\bibinfo {author} {\bibfnamefont {R.}~\bibnamefont
  {Houdr\'e}}, \bibinfo {author} {\bibfnamefont {R.~P.}\ \bibnamefont
  {Stanley}},\ and\ \bibinfo {author} {\bibfnamefont {M.}~\bibnamefont
  {Ilegems}},\ }\bibfield  {title} {\bibinfo {title} {Vacuum-field {R}abi
  splitting in the presence of inhomogeneous broadening: Resolution of a
  homogeneous linewidth in an inhomogeneously broadened system},\ }\href
  {https://doi.org/10.1103/PhysRevA.53.2711} {\bibfield  {journal} {\bibinfo
  {journal} {Phys. Rev. A}\ }\textbf {\bibinfo {volume} {53}},\ \bibinfo
  {pages} {2711} (\bibinfo {year} {1996})}\BibitemShut {NoStop}%
\bibitem [{\citenamefont {Xiang}\ \emph {et~al.}(2019)\citenamefont {Xiang},
  \citenamefont {Ribeiro}, \citenamefont {Chen}, \citenamefont {Wang},
  \citenamefont {Du}, \citenamefont {Yuen-Zhou},\ and\ \citenamefont
  {Xiong}}]{Xiang2019}%
  \BibitemOpen
  \bibfield  {author} {\bibinfo {author} {\bibfnamefont {B.}~\bibnamefont
  {Xiang}}, \bibinfo {author} {\bibfnamefont {R.~F.}\ \bibnamefont {Ribeiro}},
  \bibinfo {author} {\bibfnamefont {L.}~\bibnamefont {Chen}}, \bibinfo {author}
  {\bibfnamefont {J.}~\bibnamefont {Wang}}, \bibinfo {author} {\bibfnamefont
  {M.}~\bibnamefont {Du}}, \bibinfo {author} {\bibfnamefont {J.}~\bibnamefont
  {Yuen-Zhou}},\ and\ \bibinfo {author} {\bibfnamefont {W.}~\bibnamefont
  {Xiong}},\ }\bibfield  {title} {\bibinfo {title} {State-selective polariton
  to dark state relaxation dynamics},\ }\href
  {https://doi.org/10.1021/acs.jpca.9b04601} {\bibfield  {journal} {\bibinfo
  {journal} {J. Phys. Chem. A}\ }\textbf {\bibinfo {volume} {123}},\ \bibinfo
  {pages} {5918} (\bibinfo {year} {2019})}\BibitemShut {NoStop}%
\bibitem [{\citenamefont {Reitz}\ \emph {et~al.}(2018)\citenamefont {Reitz},
  \citenamefont {Mineo},\ and\ \citenamefont {Genes}}]{Reitz2018}%
  \BibitemOpen
  \bibfield  {author} {\bibinfo {author} {\bibfnamefont {M.}~\bibnamefont
  {Reitz}}, \bibinfo {author} {\bibfnamefont {F.}~\bibnamefont {Mineo}},\ and\
  \bibinfo {author} {\bibfnamefont {C.}~\bibnamefont {Genes}},\ }\bibfield
  {title} {\bibinfo {title} {Energy transfer and correlations in
  cavity-embedded donor-acceptor configurations},\ }\href
  {https://doi.org/10.1038/s41598-018-27396-z} {\bibfield  {journal} {\bibinfo
  {journal} {Sci. Rep.}\ }\textbf {\bibinfo {volume} {8}},\ \bibinfo {pages}
  {9050} (\bibinfo {year} {2018})}\BibitemShut {NoStop}%
\bibitem [{\citenamefont {Schäfer}\ \emph {et~al.}(2019)\citenamefont
  {Schäfer}, \citenamefont {Ruggenthaler}, \citenamefont {Appel},\ and\
  \citenamefont {Rubio}}]{Christian2019}%
  \BibitemOpen
  \bibfield  {author} {\bibinfo {author} {\bibfnamefont {C.}~\bibnamefont
  {Schäfer}}, \bibinfo {author} {\bibfnamefont {M.}~\bibnamefont
  {Ruggenthaler}}, \bibinfo {author} {\bibfnamefont {H.}~\bibnamefont
  {Appel}},\ and\ \bibinfo {author} {\bibfnamefont {A.}~\bibnamefont {Rubio}},\
  }\bibfield  {title} {\bibinfo {title} {Modification of excitation and charge
  transfer in cavity quantum-electrodynamical chemistry},\ }\href
  {https://doi.org/10.1073/pnas.1814178116} {\bibfield  {journal} {\bibinfo
  {journal} {Proc. Natl. Acad. Sci. U.S.A.}\ }\textbf {\bibinfo {volume}
  {116}},\ \bibinfo {pages} {4883} (\bibinfo {year} {2019})}\BibitemShut
  {NoStop}%
\bibitem [{\citenamefont {Cao}(2022)}]{Cao2022}%
  \BibitemOpen
  \bibfield  {author} {\bibinfo {author} {\bibfnamefont {J.}~\bibnamefont
  {Cao}},\ }\bibfield  {title} {\bibinfo {title} {Generalized resonance energy
  transfer theory: Applications to vibrational energy flow in optical
  cavities},\ }\href {https://doi.org/10.1021/acs.jpclett.2c02707} {\bibfield
  {journal} {\bibinfo  {journal} {J. Phys. Chem. Lett.}\ }\textbf {\bibinfo
  {volume} {13}},\ \bibinfo {pages} {10943} (\bibinfo {year}
  {2022})}\BibitemShut {NoStop}%
\bibitem [{\citenamefont {Cui}\ and\ \citenamefont {Nitzan}(2022)}]{Cui2022}%
  \BibitemOpen
  \bibfield  {author} {\bibinfo {author} {\bibfnamefont {B.}~\bibnamefont
  {Cui}}\ and\ \bibinfo {author} {\bibfnamefont {A.}~\bibnamefont {Nitzan}},\
  }\bibfield  {title} {\bibinfo {title} {Collective response in light-matter
  interactions: The interplay between strong coupling and local dynamics},\
  }\href@noop {} {\bibfield  {journal} {\bibinfo  {journal} {J. Chem. Phys.}\
  }\textbf {\bibinfo {volume} {157}},\ \bibinfo {pages} {114108} (\bibinfo
  {year} {2022})}\BibitemShut {NoStop}%
\bibitem [{\citenamefont {Finkelstein-Shapiro}\ \emph
  {et~al.}(2023)\citenamefont {Finkelstein-Shapiro}, \citenamefont {Mante},
  \citenamefont {Balci}, \citenamefont {Zigmantas},\ and\ \citenamefont
  {Pullerits}}]{FinkelsteinShapiro2023}%
  \BibitemOpen
  \bibfield  {author} {\bibinfo {author} {\bibfnamefont {D.}~\bibnamefont
  {Finkelstein-Shapiro}}, \bibinfo {author} {\bibfnamefont {P.-A.}\
  \bibnamefont {Mante}}, \bibinfo {author} {\bibfnamefont {S.}~\bibnamefont
  {Balci}}, \bibinfo {author} {\bibfnamefont {D.}~\bibnamefont {Zigmantas}},\
  and\ \bibinfo {author} {\bibfnamefont {T.}~\bibnamefont {Pullerits}},\
  }\bibfield  {title} {\bibinfo {title} {{Non-Hermitian Hamiltonians for linear
  and nonlinear optical response: A model for plexcitons}},\ }\href
  {https://doi.org/10.1063/5.0130287} {\bibfield  {journal} {\bibinfo
  {journal} {J. Chem. Phys.}\ }\textbf {\bibinfo {volume} {158}},\ \bibinfo
  {pages} {104104} (\bibinfo {year} {2023})}\BibitemShut {NoStop}%
\bibitem [{\citenamefont {Zhang}\ \emph {et~al.}(2023)\citenamefont {Zhang},
  \citenamefont {Nie}, \citenamefont {Lei},\ and\ \citenamefont
  {Mukamel}}]{Zhang2023}%
  \BibitemOpen
  \bibfield  {author} {\bibinfo {author} {\bibfnamefont {Z.}~\bibnamefont
  {Zhang}}, \bibinfo {author} {\bibfnamefont {X.}~\bibnamefont {Nie}}, \bibinfo
  {author} {\bibfnamefont {D.}~\bibnamefont {Lei}},\ and\ \bibinfo {author}
  {\bibfnamefont {S.}~\bibnamefont {Mukamel}},\ }\bibfield  {title} {\bibinfo
  {title} {{Multidimensional Coherent Spectroscopy of Molecular Polaritons:
  Langevin Approach}},\ }\href {https://doi.org/10.1103/PhysRevLett.130.103001}
  {\bibfield  {journal} {\bibinfo  {journal} {Phys. Rev. Lett.}\ }\textbf
  {\bibinfo {volume} {130}},\ \bibinfo {pages} {103001} (\bibinfo {year}
  {2023})}\BibitemShut {NoStop}%
\bibitem [{\citenamefont {Izrailev}\ \emph {et~al.}(1998)\citenamefont
  {Izrailev}, \citenamefont {Ruffo},\ and\ \citenamefont
  {Tessieri}}]{Izrailev1998}%
  \BibitemOpen
  \bibfield  {author} {\bibinfo {author} {\bibfnamefont {F.~M.}\ \bibnamefont
  {Izrailev}}, \bibinfo {author} {\bibfnamefont {S.}~\bibnamefont {Ruffo}},\
  and\ \bibinfo {author} {\bibfnamefont {L.}~\bibnamefont {Tessieri}},\
  }\bibfield  {title} {\bibinfo {title} {Classical representation of the
  one-dimensional {Anderson} model},\ }\href
  {https://doi.org/10.1088/0305-4470/31/23/008} {\bibfield  {journal} {\bibinfo
   {journal} {J. Phys. A}\ }\textbf {\bibinfo {volume} {31}},\ \bibinfo {pages}
  {5263} (\bibinfo {year} {1998})}\BibitemShut {NoStop}%
\bibitem [{\citenamefont {Agranovich}\ \emph {et~al.}(2003)\citenamefont
  {Agranovich}, \citenamefont {Litinskaia},\ and\ \citenamefont
  {Lidzey}}]{Agranovich2003}%
  \BibitemOpen
  \bibfield  {author} {\bibinfo {author} {\bibfnamefont {V.~M.}\ \bibnamefont
  {Agranovich}}, \bibinfo {author} {\bibfnamefont {M.}~\bibnamefont
  {Litinskaia}},\ and\ \bibinfo {author} {\bibfnamefont {D.~G.}\ \bibnamefont
  {Lidzey}},\ }\bibfield  {title} {\bibinfo {title} {Cavity polaritons in
  microcavities containing disordered organic semiconductors},\ }\href
  {https://doi.org/10.1103/PhysRevB.67.085311} {\bibfield  {journal} {\bibinfo
  {journal} {Phys. Rev. B}\ }\textbf {\bibinfo {volume} {67}},\ \bibinfo
  {pages} {085311} (\bibinfo {year} {2003})}\BibitemShut {NoStop}%
\bibitem [{\citenamefont {Litinskaya}\ and\ \citenamefont
  {Reineker}(2006)}]{Litinskaya2006}%
  \BibitemOpen
  \bibfield  {author} {\bibinfo {author} {\bibfnamefont {M.}~\bibnamefont
  {Litinskaya}}\ and\ \bibinfo {author} {\bibfnamefont {P.}~\bibnamefont
  {Reineker}},\ }\bibfield  {title} {\bibinfo {title} {Loss of coherence of
  exciton polaritons in inhomogeneous organic microcavities},\ }\href
  {https://doi.org/10.1103/PhysRevB.74.165320} {\bibfield  {journal} {\bibinfo
  {journal} {Phys. Rev. B}\ }\textbf {\bibinfo {volume} {74}},\ \bibinfo
  {pages} {165320} (\bibinfo {year} {2006})}\BibitemShut {NoStop}%
\bibitem [{\citenamefont {Litinskaya}\ \emph {et~al.}(2004)\citenamefont
  {Litinskaya}, \citenamefont {Reineker},\ and\ \citenamefont
  {Agranovich}}]{Litinskaya2004}%
  \BibitemOpen
  \bibfield  {author} {\bibinfo {author} {\bibfnamefont {M.}~\bibnamefont
  {Litinskaya}}, \bibinfo {author} {\bibfnamefont {P.}~\bibnamefont
  {Reineker}},\ and\ \bibinfo {author} {\bibfnamefont {V.}~\bibnamefont
  {Agranovich}},\ }\bibfield  {title} {\bibinfo {title} {Fast polariton
  relaxation in strongly coupled organic microcavities},\ }\href
  {https://doi.org/https://doi.org/10.1016/j.jlumin.2004.08.033} {\bibfield
  {journal} {\bibinfo  {journal} {J. Lumin.}\ }\textbf {\bibinfo {volume}
  {110}},\ \bibinfo {pages} {364} (\bibinfo {year} {2004})}\BibitemShut
  {NoStop}%
\bibitem [{\citenamefont {Ribeiro}(2022)}]{Ribeiro2022}%
  \BibitemOpen
  \bibfield  {author} {\bibinfo {author} {\bibfnamefont {R.~F.}\ \bibnamefont
  {Ribeiro}},\ }\bibfield  {title} {\bibinfo {title} {Multimode polariton
  effects on molecular energy transport and spectral fluctuations},\ }\href
  {https://doi.org/10.1038/s42004-022-00660-0} {\bibfield  {journal} {\bibinfo
  {journal} {Comm. Chem.}\ }\textbf {\bibinfo {volume} {5}},\ \bibinfo {pages}
  {48} (\bibinfo {year} {2022})}\BibitemShut {NoStop}%
\bibitem [{\citenamefont {Allard}\ and\ \citenamefont
  {Weick}(2022)}]{Allard2022}%
  \BibitemOpen
  \bibfield  {author} {\bibinfo {author} {\bibfnamefont {T.~F.}\ \bibnamefont
  {Allard}}\ and\ \bibinfo {author} {\bibfnamefont {G.}~\bibnamefont {Weick}},\
  }\bibfield  {title} {\bibinfo {title} {Disorder-enhanced transport in a chain
  of lossy dipoles strongly coupled to cavity photons},\ }\href
  {https://doi.org/10.1103/PhysRevB.106.245424} {\bibfield  {journal} {\bibinfo
   {journal} {Phys. Rev. B}\ }\textbf {\bibinfo {volume} {106}},\ \bibinfo
  {pages} {245424} (\bibinfo {year} {2022})}\BibitemShut {NoStop}%
\bibitem [{\citenamefont {Patton}\ \emph {et~al.}()\citenamefont {Patton},
  \citenamefont {Norman}, \citenamefont {Scalettar},\ and\ \citenamefont
  {Radulaski}}]{Patton2021}%
  \BibitemOpen
  \bibfield  {author} {\bibinfo {author} {\bibfnamefont {J.}~\bibnamefont
  {Patton}}, \bibinfo {author} {\bibfnamefont {V.}~\bibnamefont {Norman}},
  \bibinfo {author} {\bibfnamefont {R.}~\bibnamefont {Scalettar}},\ and\
  \bibinfo {author} {\bibfnamefont {M.}~\bibnamefont {Radulaski}},\ }\href@noop
  {} {\bibinfo {title} {All-photonic quantum simulators with spectrally
  disordered emitters}},\ \bibinfo {note} {arXiv:2112.15469}\BibitemShut
  {NoStop}%
\bibitem [{\citenamefont {Ćwik}\ \emph {et~al.}(2014)\citenamefont {Ćwik},
  \citenamefont {Reja}, \citenamefont {Littlewood},\ and\ \citenamefont
  {Keeling}}]{Cwik2014}%
  \BibitemOpen
  \bibfield  {author} {\bibinfo {author} {\bibfnamefont {J.~A.}\ \bibnamefont
  {Ćwik}}, \bibinfo {author} {\bibfnamefont {S.}~\bibnamefont {Reja}},
  \bibinfo {author} {\bibfnamefont {P.~B.}\ \bibnamefont {Littlewood}},\ and\
  \bibinfo {author} {\bibfnamefont {J.}~\bibnamefont {Keeling}},\ }\bibfield
  {title} {\bibinfo {title} {Polariton condensation with saturable molecules
  dressed by vibrational modes},\ }\href
  {https://doi.org/10.1209/0295-5075/105/47009} {\bibfield  {journal} {\bibinfo
   {journal} {Europhys. Lett.}\ }\textbf {\bibinfo {volume} {105}},\ \bibinfo
  {pages} {47009} (\bibinfo {year} {2014})}\BibitemShut {NoStop}%
\bibitem [{\citenamefont {Strashko}\ \emph {et~al.}(2018)\citenamefont
  {Strashko}, \citenamefont {Kirton},\ and\ \citenamefont
  {Keeling}}]{Strashko2018}%
  \BibitemOpen
  \bibfield  {author} {\bibinfo {author} {\bibfnamefont {A.}~\bibnamefont
  {Strashko}}, \bibinfo {author} {\bibfnamefont {P.}~\bibnamefont {Kirton}},\
  and\ \bibinfo {author} {\bibfnamefont {J.}~\bibnamefont {Keeling}},\
  }\bibfield  {title} {\bibinfo {title} {{Organic Polariton Lasing and the Weak
  to Strong Coupling Crossover}},\ }\href
  {https://doi.org/10.1103/PhysRevLett.121.193601} {\bibfield  {journal}
  {\bibinfo  {journal} {Phys. Rev. Lett.}\ }\textbf {\bibinfo {volume} {121}},\
  \bibinfo {pages} {193601} (\bibinfo {year} {2018})}\BibitemShut {NoStop}%
\bibitem [{\citenamefont {Sokolovskii}\ \emph {et~al.}()\citenamefont
  {Sokolovskii}, \citenamefont {Tichauer}, \citenamefont {Feist},\ and\
  \citenamefont {Groenhof}}]{Sokolovskii2022}%
  \BibitemOpen
  \bibfield  {author} {\bibinfo {author} {\bibfnamefont {I.}~\bibnamefont
  {Sokolovskii}}, \bibinfo {author} {\bibfnamefont {R.~H.}\ \bibnamefont
  {Tichauer}}, \bibinfo {author} {\bibfnamefont {J.}~\bibnamefont {Feist}},\
  and\ \bibinfo {author} {\bibfnamefont {G.}~\bibnamefont {Groenhof}},\
  }\href@noop {} {\bibinfo {title} {Enhanced excitation energy transfer under
  strong light-matter coupling: Insights from multi-scale molecular dynamics
  simulations}},\ \Eprint {https://arxiv.org/abs/2209.07309} {arXiv:2209.07309}
  \BibitemShut {NoStop}%
\bibitem [{\citenamefont {Xu}\ \emph {et~al.}()\citenamefont {Xu},
  \citenamefont {Mandal}, \citenamefont {Baxter}, \citenamefont {Cheng},
  \citenamefont {Lee}, \citenamefont {Su}, \citenamefont {Liu}, \citenamefont
  {Reichman},\ and\ \citenamefont {Delor}}]{Xu2022}%
  \BibitemOpen
  \bibfield  {author} {\bibinfo {author} {\bibfnamefont {D.}~\bibnamefont
  {Xu}}, \bibinfo {author} {\bibfnamefont {A.}~\bibnamefont {Mandal}}, \bibinfo
  {author} {\bibfnamefont {J.~M.}\ \bibnamefont {Baxter}}, \bibinfo {author}
  {\bibfnamefont {S.~W.}\ \bibnamefont {Cheng}}, \bibinfo {author}
  {\bibfnamefont {I.}~\bibnamefont {Lee}}, \bibinfo {author} {\bibfnamefont
  {H.}~\bibnamefont {Su}}, \bibinfo {author} {\bibfnamefont {S.}~\bibnamefont
  {Liu}}, \bibinfo {author} {\bibfnamefont {D.~R.}\ \bibnamefont {Reichman}},\
  and\ \bibinfo {author} {\bibfnamefont {M.}~\bibnamefont {Delor}},\
  }\href@noop {} {\bibinfo {title} {Ultrafast imaging of coherent polariton
  propagation and interactions}},\ \bibinfo {note}
  {arXiv:2205.01176}\BibitemShut {NoStop}%
\bibitem [{\citenamefont {Alvertis}\ \emph {et~al.}(2020)\citenamefont
  {Alvertis}, \citenamefont {Pandya}, \citenamefont {Quarti}, \citenamefont
  {Legrand}, \citenamefont {Barisien}, \citenamefont {Monserrat}, \citenamefont
  {Musser}, \citenamefont {Rao}, \citenamefont {Chin},\ and\ \citenamefont
  {Beljonne}}]{Alvertis2020}%
  \BibitemOpen
  \bibfield  {author} {\bibinfo {author} {\bibfnamefont {A.~M.}\ \bibnamefont
  {Alvertis}}, \bibinfo {author} {\bibfnamefont {R.}~\bibnamefont {Pandya}},
  \bibinfo {author} {\bibfnamefont {C.}~\bibnamefont {Quarti}}, \bibinfo
  {author} {\bibfnamefont {L.}~\bibnamefont {Legrand}}, \bibinfo {author}
  {\bibfnamefont {T.}~\bibnamefont {Barisien}}, \bibinfo {author}
  {\bibfnamefont {B.}~\bibnamefont {Monserrat}}, \bibinfo {author}
  {\bibfnamefont {A.~J.}\ \bibnamefont {Musser}}, \bibinfo {author}
  {\bibfnamefont {A.}~\bibnamefont {Rao}}, \bibinfo {author} {\bibfnamefont
  {A.~W.}\ \bibnamefont {Chin}},\ and\ \bibinfo {author} {\bibfnamefont
  {D.}~\bibnamefont {Beljonne}},\ }\bibfield  {title} {\bibinfo {title} {First
  principles modeling of exciton-polaritons in polydiacetylene chains},\ }\href
  {https://doi.org/10.1063/5.0019009} {\bibfield  {journal} {\bibinfo
  {journal} {J. Chem. Phys.}\ }\textbf {\bibinfo {volume} {153}},\ \bibinfo
  {pages} {084103} (\bibinfo {year} {2020})}\BibitemShut {NoStop}%
\bibitem [{\citenamefont {Anderson}(1958)}]{Anderson1958}%
  \BibitemOpen
  \bibfield  {author} {\bibinfo {author} {\bibfnamefont {P.~W.}\ \bibnamefont
  {Anderson}},\ }\bibfield  {title} {\bibinfo {title} {Absence of diffusion in
  certain random lattices},\ }\href {https://doi.org/10.1103/PhysRev.109.1492}
  {\bibfield  {journal} {\bibinfo  {journal} {Phys. Rev.}\ }\textbf {\bibinfo
  {volume} {109}},\ \bibinfo {pages} {1492} (\bibinfo {year}
  {1958})}\BibitemShut {NoStop}%
\bibitem [{\citenamefont {Abrahams}\ \emph {et~al.}(1979)\citenamefont
  {Abrahams}, \citenamefont {Anderson}, \citenamefont {Licciardello},\ and\
  \citenamefont {Ramakrishnan}}]{Abrahams1979}%
  \BibitemOpen
  \bibfield  {author} {\bibinfo {author} {\bibfnamefont {E.}~\bibnamefont
  {Abrahams}}, \bibinfo {author} {\bibfnamefont {P.~W.}\ \bibnamefont
  {Anderson}}, \bibinfo {author} {\bibfnamefont {D.~C.}\ \bibnamefont
  {Licciardello}},\ and\ \bibinfo {author} {\bibfnamefont {T.~V.}\ \bibnamefont
  {Ramakrishnan}},\ }\bibfield  {title} {\bibinfo {title} {{Scaling Theory of
  Localization: Absence of Quantum Diffusion in Two Dimensions}},\ }\href
  {https://doi.org/10.1103/PhysRevLett.42.673} {\bibfield  {journal} {\bibinfo
  {journal} {Phys. Rev. Lett.}\ }\textbf {\bibinfo {volume} {42}},\ \bibinfo
  {pages} {673} (\bibinfo {year} {1979})}\BibitemShut {NoStop}%
\bibitem [{\citenamefont {Wang}\ \emph {et~al.}(2020)\citenamefont {Wang},
  \citenamefont {Xia}, \citenamefont {Zhang}, \citenamefont {Yao},
  \citenamefont {Chen}, \citenamefont {You}, \citenamefont {Zhou},\ and\
  \citenamefont {Liu}}]{Wang2020}%
  \BibitemOpen
  \bibfield  {author} {\bibinfo {author} {\bibfnamefont {Y.}~\bibnamefont
  {Wang}}, \bibinfo {author} {\bibfnamefont {X.}~\bibnamefont {Xia}}, \bibinfo
  {author} {\bibfnamefont {L.}~\bibnamefont {Zhang}}, \bibinfo {author}
  {\bibfnamefont {H.}~\bibnamefont {Yao}}, \bibinfo {author} {\bibfnamefont
  {S.}~\bibnamefont {Chen}}, \bibinfo {author} {\bibfnamefont {J.}~\bibnamefont
  {You}}, \bibinfo {author} {\bibfnamefont {Q.}~\bibnamefont {Zhou}},\ and\
  \bibinfo {author} {\bibfnamefont {X.-J.}\ \bibnamefont {Liu}},\ }\bibfield
  {title} {\bibinfo {title} {{One-Dimensional Quasiperiodic Mosaic Lattice with
  Exact Mobility Edges}},\ }\href
  {https://doi.org/10.1103/PhysRevLett.125.196604} {\bibfield  {journal}
  {\bibinfo  {journal} {Phys. Rev. Lett.}\ }\textbf {\bibinfo {volume} {125}},\
  \bibinfo {pages} {196604} (\bibinfo {year} {2020})}\BibitemShut {NoStop}%
\bibitem [{\citenamefont {Cao}\ and\ \citenamefont {Silbey}(2009)}]{Cao2009}%
  \BibitemOpen
  \bibfield  {author} {\bibinfo {author} {\bibfnamefont {J.}~\bibnamefont
  {Cao}}\ and\ \bibinfo {author} {\bibfnamefont {R.~J.}\ \bibnamefont
  {Silbey}},\ }\bibfield  {title} {\bibinfo {title} {Optimization of exciton
  trapping in energy transfer processes},\ }\href@noop {} {\bibfield  {journal}
  {\bibinfo  {journal} {J. Phys. Chem. A}\ }\textbf {\bibinfo {volume} {113}},\
  \bibinfo {pages} {13825} (\bibinfo {year} {2009})}\BibitemShut {NoStop}%
\bibitem [{\citenamefont {Chuang}\ \emph {et~al.}(2016)\citenamefont {Chuang},
  \citenamefont {Lee}, \citenamefont {Moix}, \citenamefont {Knoester},\ and\
  \citenamefont {Cao}}]{Chuang2016}%
  \BibitemOpen
  \bibfield  {author} {\bibinfo {author} {\bibfnamefont {C.}~\bibnamefont
  {Chuang}}, \bibinfo {author} {\bibfnamefont {C.~K.}\ \bibnamefont {Lee}},
  \bibinfo {author} {\bibfnamefont {J.~M.}\ \bibnamefont {Moix}}, \bibinfo
  {author} {\bibfnamefont {J.}~\bibnamefont {Knoester}},\ and\ \bibinfo
  {author} {\bibfnamefont {J.}~\bibnamefont {Cao}},\ }\bibfield  {title}
  {\bibinfo {title} {{Quantum Diffusion on Molecular Tubes: Universal Scaling
  of the 1D to 2D Transition}},\ }\href@noop {} {\bibfield  {journal} {\bibinfo
   {journal} {Phys. Rev. Lett.}\ }\textbf {\bibinfo {volume} {116}},\ \bibinfo
  {pages} {196803} (\bibinfo {year} {2016})}\BibitemShut {NoStop}%
\bibitem [{sup()}]{supplementals}%
  \BibitemOpen
  \bibinfo {note} {See Supplementary Material for more information about the
  solution of the TC model and the numerical calculations.}\BibitemShut {Stop}%
\bibitem [{\citenamefont {Engelhardt}\ \emph {et~al.}(2016)\citenamefont
  {Engelhardt}, \citenamefont {Schaller},\ and\ \citenamefont
  {Brandes}}]{Engelhardt2016a}%
  \BibitemOpen
  \bibfield  {author} {\bibinfo {author} {\bibfnamefont {G.}~\bibnamefont
  {Engelhardt}}, \bibinfo {author} {\bibfnamefont {G.}~\bibnamefont
  {Schaller}},\ and\ \bibinfo {author} {\bibfnamefont {T.}~\bibnamefont
  {Brandes}},\ }\bibfield  {title} {\bibinfo {title} {{Bosonic Josephson effect
  in the Fano-Anderson model}},\ }\href
  {https://doi.org/10.1103/PhysRevA.94.013608} {\bibfield  {journal} {\bibinfo
  {journal} {Phys. Rev. A}\ }\textbf {\bibinfo {volume} {94}},\ \bibinfo
  {pages} {013608} (\bibinfo {year} {2016})}\BibitemShut {NoStop}%
\bibitem [{\citenamefont {Topp}\ \emph {et~al.}(2015)\citenamefont {Topp},
  \citenamefont {Schaller},\ and\ \citenamefont {Brandes}}]{Topp2015}%
  \BibitemOpen
  \bibfield  {author} {\bibinfo {author} {\bibfnamefont {G.~E.}\ \bibnamefont
  {Topp}}, \bibinfo {author} {\bibfnamefont {G.}~\bibnamefont {Schaller}},\
  and\ \bibinfo {author} {\bibfnamefont {T.}~\bibnamefont {Brandes}},\
  }\bibfield  {title} {\bibinfo {title} {Steady-state thermodynamics of
  non-interacting transport beyond weak coupling},\ }\href@noop {} {\bibfield
  {journal} {\bibinfo  {journal} {Europhys. Lett.}\ }\textbf {\bibinfo {volume}
  {110}},\ \bibinfo {pages} {67003} (\bibinfo {year} {2015})}\BibitemShut
  {NoStop}%
\bibitem [{\citenamefont {Fleischhauer}\ \emph {et~al.}(2005)\citenamefont
  {Fleischhauer}, \citenamefont {Imamoglu},\ and\ \citenamefont
  {Marangos}}]{Fleischhauer2005}%
  \BibitemOpen
  \bibfield  {author} {\bibinfo {author} {\bibfnamefont {M.}~\bibnamefont
  {Fleischhauer}}, \bibinfo {author} {\bibfnamefont {A.}~\bibnamefont
  {Imamoglu}},\ and\ \bibinfo {author} {\bibfnamefont {J.~P.}\ \bibnamefont
  {Marangos}},\ }\bibfield  {title} {\bibinfo {title} {Electromagnetically
  induced transparency: Optics in coherent media},\ }\href
  {https://doi.org/10.1103/RevModPhys.77.633} {\bibfield  {journal} {\bibinfo
  {journal} {Rev. Mod. Phys.}\ }\textbf {\bibinfo {volume} {77}},\ \bibinfo
  {pages} {633} (\bibinfo {year} {2005})}\BibitemShut {NoStop}%
\bibitem [{\citenamefont {Engelhardt}\ and\ \citenamefont
  {Cao}(2021)}]{Engelhardt2021}%
  \BibitemOpen
  \bibfield  {author} {\bibinfo {author} {\bibfnamefont {G.}~\bibnamefont
  {Engelhardt}}\ and\ \bibinfo {author} {\bibfnamefont {J.}~\bibnamefont
  {Cao}},\ }\bibfield  {title} {\bibinfo {title} {{Dynamical Symmetries and
  Symmetry-Protected Selection Rules in Periodically Driven Quantum Systems}},\
  }\href {https://doi.org/10.1103/PhysRevLett.126.090601} {\bibfield  {journal}
  {\bibinfo  {journal} {Phys. Rev. Lett.}\ }\textbf {\bibinfo {volume} {126}},\
  \bibinfo {pages} {090601} (\bibinfo {year} {2021})}\BibitemShut {NoStop}%
\bibitem [{\citenamefont {Herrera}\ and\ \citenamefont
  {Litinskaya}(2022)}]{Herrera2022}%
  \BibitemOpen
  \bibfield  {author} {\bibinfo {author} {\bibfnamefont {F.}~\bibnamefont
  {Herrera}}\ and\ \bibinfo {author} {\bibfnamefont {M.}~\bibnamefont
  {Litinskaya}},\ }\bibfield  {title} {\bibinfo {title} {Disordered ensembles
  of strongly coupled single-molecule plasmonic picocavities as nonlinear
  optical metamaterials},\ }\href {https://doi.org/10.1063/5.0080063}
  {\bibfield  {journal} {\bibinfo  {journal} {J. Chem. Phys.}\ }\textbf
  {\bibinfo {volume} {156}},\ \bibinfo {pages} {114702} (\bibinfo {year}
  {2022})}\BibitemShut {NoStop}%
\bibitem [{\citenamefont {Reed}\ and\ \citenamefont {Barry}(1975)}]{Reed1975}%
  \BibitemOpen
  \bibfield  {author} {\bibinfo {author} {\bibfnamefont {M.}~\bibnamefont
  {Reed}}\ and\ \bibinfo {author} {\bibfnamefont {S.}~\bibnamefont {Barry}},\
  }\href@noop {} {\emph {\bibinfo {title} {{Methods of Modern Mathematical
  Physics. II. Fourier Analysis, Self-Adjointness.}}}}\ (\bibinfo  {publisher}
  {Academic Press, New-York-London},\ \bibinfo {year} {1975})\ \bibinfo {note}
  {{Section} IX.3; Theorem IX.13.}\BibitemShut {Stop}%
\bibitem [{\citenamefont {Wu}\ \emph {et~al.}(2013)\citenamefont {Wu},
  \citenamefont {Silbey},\ and\ \citenamefont {Cao}}]{Wu2013}%
  \BibitemOpen
  \bibfield  {author} {\bibinfo {author} {\bibfnamefont {J.}~\bibnamefont
  {Wu}}, \bibinfo {author} {\bibfnamefont {R.~J.}\ \bibnamefont {Silbey}},\
  and\ \bibinfo {author} {\bibfnamefont {J.}~\bibnamefont {Cao}},\ }\bibfield
  {title} {\bibinfo {title} {{Generic Mechanism of Optimal Energy Transfer
  Efficiency: A Scaling Theory of the Mean First-Passage Time in Exciton
  Systems}},\ }\href@noop {} {\bibfield  {journal} {\bibinfo  {journal} {Phys.
  Rev. Lett.}\ }\textbf {\bibinfo {volume} {110}},\ \bibinfo {pages} {200402}
  (\bibinfo {year} {2013})}\BibitemShut {NoStop}%
\bibitem [{\citenamefont {Lee}\ \emph {et~al.}(2015)\citenamefont {Lee},
  \citenamefont {Moix},\ and\ \citenamefont {Cao}}]{Lee2015}%
  \BibitemOpen
  \bibfield  {author} {\bibinfo {author} {\bibfnamefont {C.~K.}\ \bibnamefont
  {Lee}}, \bibinfo {author} {\bibfnamefont {J.}~\bibnamefont {Moix}},\ and\
  \bibinfo {author} {\bibfnamefont {J.}~\bibnamefont {Cao}},\ }\bibfield
  {title} {\bibinfo {title} {Coherent quantum transport in disordered systems:
  A unified polaron treatment of hopping and band-like transport},\ }\href@noop
  {} {\bibfield  {journal} {\bibinfo  {journal} {J. Chem. Phys.}\ }\textbf
  {\bibinfo {volume} {142}},\ \bibinfo {pages} {164103} (\bibinfo {year}
  {2015})}\BibitemShut {NoStop}%
\bibitem [{\citenamefont {Engelhardt}\ and\ \citenamefont
  {Cao}(2019)}]{Engelhardt2019a}%
  \BibitemOpen
  \bibfield  {author} {\bibinfo {author} {\bibfnamefont {G.}~\bibnamefont
  {Engelhardt}}\ and\ \bibinfo {author} {\bibfnamefont {J.}~\bibnamefont
  {Cao}},\ }\bibfield  {title} {\bibinfo {title} {Tuning the {Aharonov-Bohm}
  effect with dephasing in nonequilibrium transport},\ }\href
  {https://doi.org/10.1103/PhysRevB.99.075436} {\bibfield  {journal} {\bibinfo
  {journal} {Phys. Rev. B}\ }\textbf {\bibinfo {volume} {99}},\ \bibinfo
  {pages} {075436} (\bibinfo {year} {2019})}\BibitemShut {NoStop}%
\bibitem [{\citenamefont {Chenu}\ and\ \citenamefont {Cao}(2017)}]{Chenu2017}%
  \BibitemOpen
  \bibfield  {author} {\bibinfo {author} {\bibfnamefont {A.}~\bibnamefont
  {Chenu}}\ and\ \bibinfo {author} {\bibfnamefont {J.}~\bibnamefont {Cao}},\
  }\bibfield  {title} {\bibinfo {title} {{Construction of Multichromophoric
  Spectra from Monomer Data: Applications to Resonant Energy Transfer}},\
  }\href@noop {} {\bibfield  {journal} {\bibinfo  {journal} {Phys. Rev. Lett.}\
  }\textbf {\bibinfo {volume} {118}},\ \bibinfo {pages} {013001} (\bibinfo
  {year} {2017})}\BibitemShut {NoStop}%
\bibitem [{\citenamefont {Pandya}\ \emph {et~al.}(2022)\citenamefont {Pandya},
  \citenamefont {Ashoka}, \citenamefont {Georgiou}, \citenamefont {Sung},
  \citenamefont {Jayaprakash}, \citenamefont {Renken}, \citenamefont {Gai},
  \citenamefont {Shen}, \citenamefont {Rao},\ and\ \citenamefont
  {Musser}}]{Pandya2022}%
  \BibitemOpen
  \bibfield  {author} {\bibinfo {author} {\bibfnamefont {R.}~\bibnamefont
  {Pandya}}, \bibinfo {author} {\bibfnamefont {A.}~\bibnamefont {Ashoka}},
  \bibinfo {author} {\bibfnamefont {K.}~\bibnamefont {Georgiou}}, \bibinfo
  {author} {\bibfnamefont {J.}~\bibnamefont {Sung}}, \bibinfo {author}
  {\bibfnamefont {R.}~\bibnamefont {Jayaprakash}}, \bibinfo {author}
  {\bibfnamefont {S.}~\bibnamefont {Renken}}, \bibinfo {author} {\bibfnamefont
  {L.}~\bibnamefont {Gai}}, \bibinfo {author} {\bibfnamefont {Z.}~\bibnamefont
  {Shen}}, \bibinfo {author} {\bibfnamefont {A.}~\bibnamefont {Rao}},\ and\
  \bibinfo {author} {\bibfnamefont {A.~J.}\ \bibnamefont {Musser}},\ }\bibfield
   {title} {\bibinfo {title} {Tuning the coherent propagation of organic
  exciton-polaritons through dark state delocalization},\ }\href
  {https://doi.org/https://doi.org/10.1002/advs.202105569} {\bibfield
  {journal} {\bibinfo  {journal} {Adv. Sci.}\ }\textbf {\bibinfo {volume}
  {9}},\ \bibinfo {pages} {2105569} (\bibinfo {year} {2022})}\BibitemShut
  {NoStop}%
\end{thebibliography}%

\begin{widetext}
\newpage 
\begin{center}
	\Huge
	\textbf{Supplementary Materials}
\end{center}

\end{widetext}

\setcounter{equation}{0}
\setcounter{figure}{0}

\section{ Analytical calculations}

\label{sec:GreensFct}

\subsection{System}

We model an ensemble of disordered quantum emitters in a microcavity by the  multimode disordered Tavis-Cummings Hamiltonian
\begin{equation}
\hat H = \hat H_{\text{M}} + \hat H_{\text{L}} + \hat H_{\text{LM}},
\label{eq:Hamiltonian}
\end{equation}
where
\begin{eqnarray}
\hat H_{\text{M}} &=& \sum_{j=1}^{N} E_j \hat B_{j}^\dagger \hat B_j  \nonumber, \qquad
\hat H_{\text{L}} = \sum_{k} \;  \omega_{ k}   \hat a_{ k }^\dagger\hat a_{ k}  ,   \nonumber  \\   
\hat H_{\text{LM}} &=&   \sum_{j=1}^{N}\sum_{ k}      g_{j,k} \hat B_{j}^\dagger    \hat  a_{ k}  + \text{H.c.}  \quad 
\label{eq:hamiltoniancontinuumModeTavisCummingsModel}
\end{eqnarray}
Here,  $j$ and $k$ label the quantum emitters $\hat B_j $ and the photonic modes $\hat a_k$, respectively, both fulfilling a bosonic commutation relation. The quantum emitters can represent atoms, molecules or particle-hole pairs. For concreteness, we specify to molecules in this work.  The excitation energies of the molecules $E_j$ are distributed according to a probability distribution $P(E)$. In this work, we mainly consider a Gaussian distribution, 
\begin{equation}
P (E ) =\frac{1}{\sqrt{\pi} \sigma} e^{-\frac{(E-E_{\text{M} }) ^2}{2\sigma^2} }
\end{equation}
with center $E_{\text{M} }$ and width $\sigma$, but our findings are also valid for arbitrary disorder models.  We consider a one-dimensional translational-invariant system of length $L$ with a periodic boundary condition. The molecules are located at positions $r_j = j L/N $. The photonic dispersion relation is given by
\begin{equation}
\omega_k = \sqrt{c^2 q_k^2 + E_{C }^2},
\label{eq:photonDispRelation}
\end{equation}
with $q_k = 2\pi k/L $ and integer $k$. The confinement energy $E_{C }$ appears because of the spatial confinement of the light field in the transversal direction of the microcavity. 
The photonic mode functions are given by $\varphi_k(r) = e^{i q_k r}/\sqrt{L}$. The light-matter coupling is parameterized by $g_{j,k} =  g_k  \varphi_k(r_j)$. As the excitation number $\sum_k \hat a_k ^\dagger \hat a_k + \sum_j\hat B_{j}^\dagger \hat B_j  $ is an integral of motion, we focus on the single-excitation manifold for simplicity. In this work, we derive a closed-form expression for  the Green's function of the multimode Tavis-Cummings model in the thermodynamic limit, that we define by $N,L\rightarrow \infty$ for a constant molecule density $\rho =N/L$.

\subsection{No disorder}

\label{sec:noDisorder}

\begin{figure}[t]
	\includegraphics[width=\linewidth]{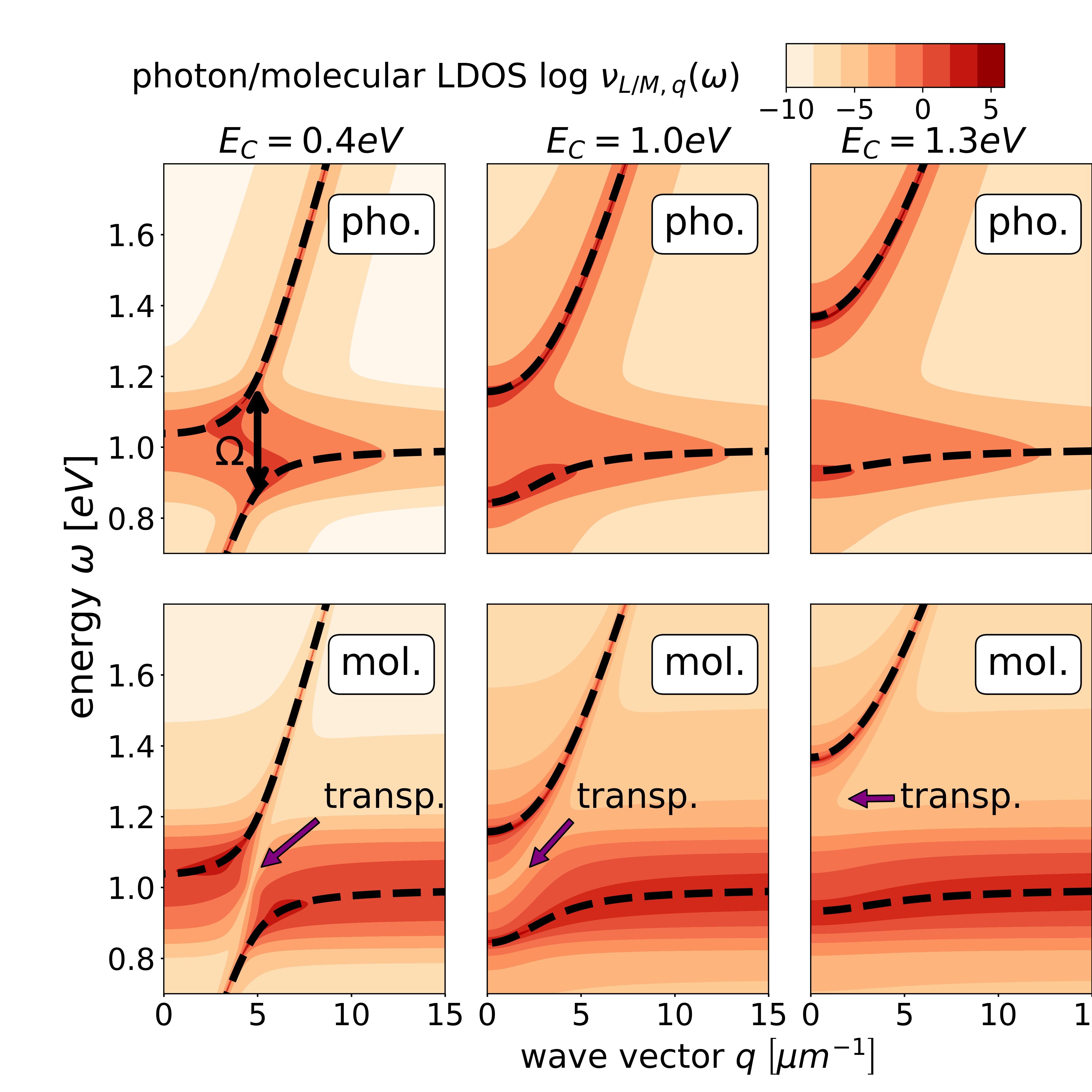}
	\caption{ Wave-vector-resolved  photon and molecule LDOSs for $E_{C }=0.4 \,\text{eV}$,   $E_{C }=1.0\,\text{eV}$ and $E_{C }=1.3\,\text{eV}$.   Overall parameters are  $L = 124\,\mu \text{m}$,   $N= 5000$, $E_{\text{M} } = 1.0\,\text{eV}$, $\sigma = 0.05\,\text{eV}$, and $g\sqrt{\rho}= 0.14\,\text{eV}$.  }
	\label{figLocalDensityOfStates}
\end{figure}

\begin{figure}[t]
	\includegraphics[width=\linewidth]{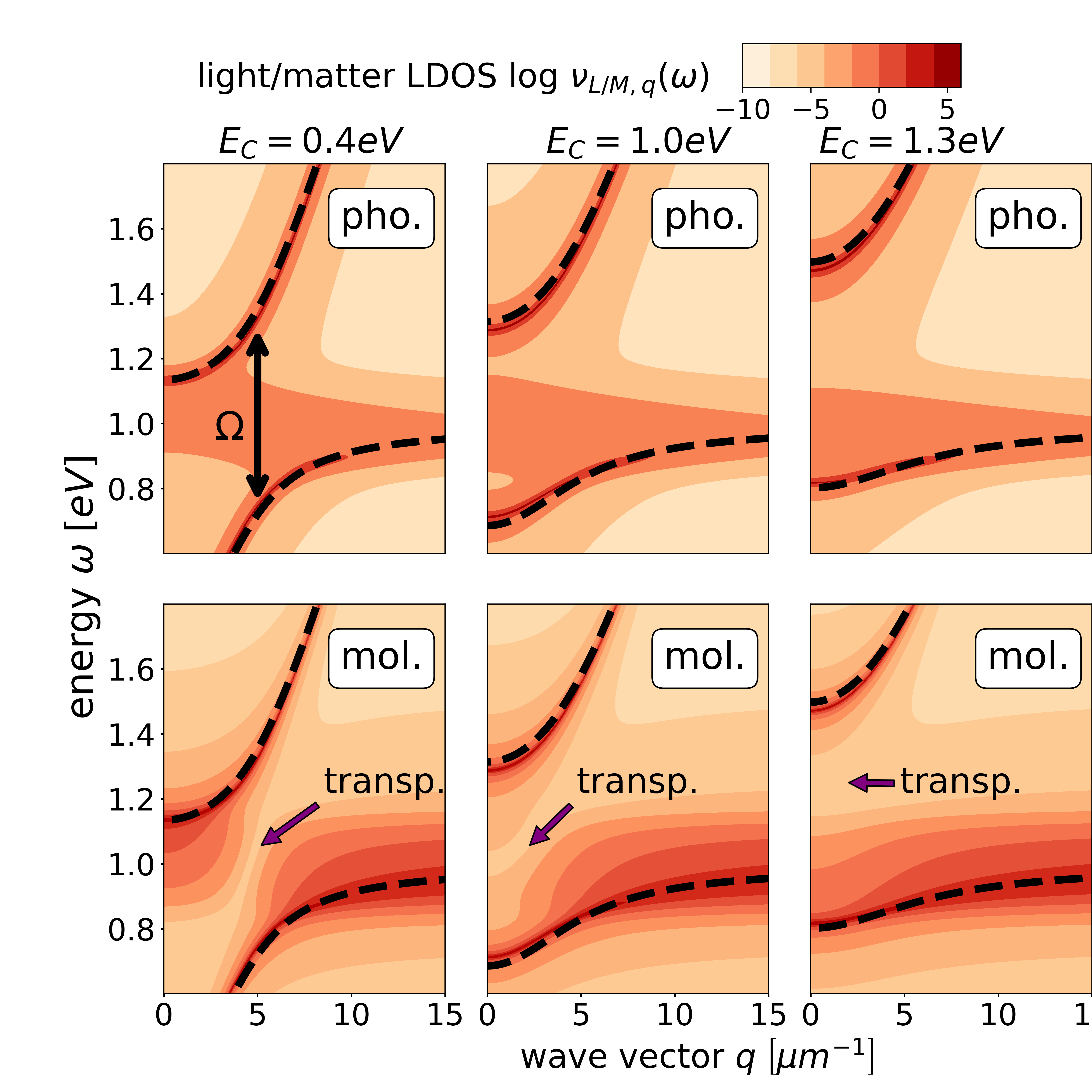}
	\caption{ Wave-vector-resolved  photon and molecule LDOSs for $E_{C }=0.4 \,\text{eV}$,  $E_{C }=1.0\,\text{eV}$ and $E_{C }=1.3\,\text{eV}$.   Overall parameters are  $L = 124\,\mu \text{m}$,   $N= 5000$, $E_{\text{M} } = 1.0\,\text{eV}$, $\sigma = 0.05\,\text{eV}$, and $g\sqrt{\rho} = 0.28\,\text{eV}$.  }
	\label{figLocalDensityOfStatesLMintLarge}
\end{figure}

\begin{figure}[t]
	\includegraphics[width=\linewidth]{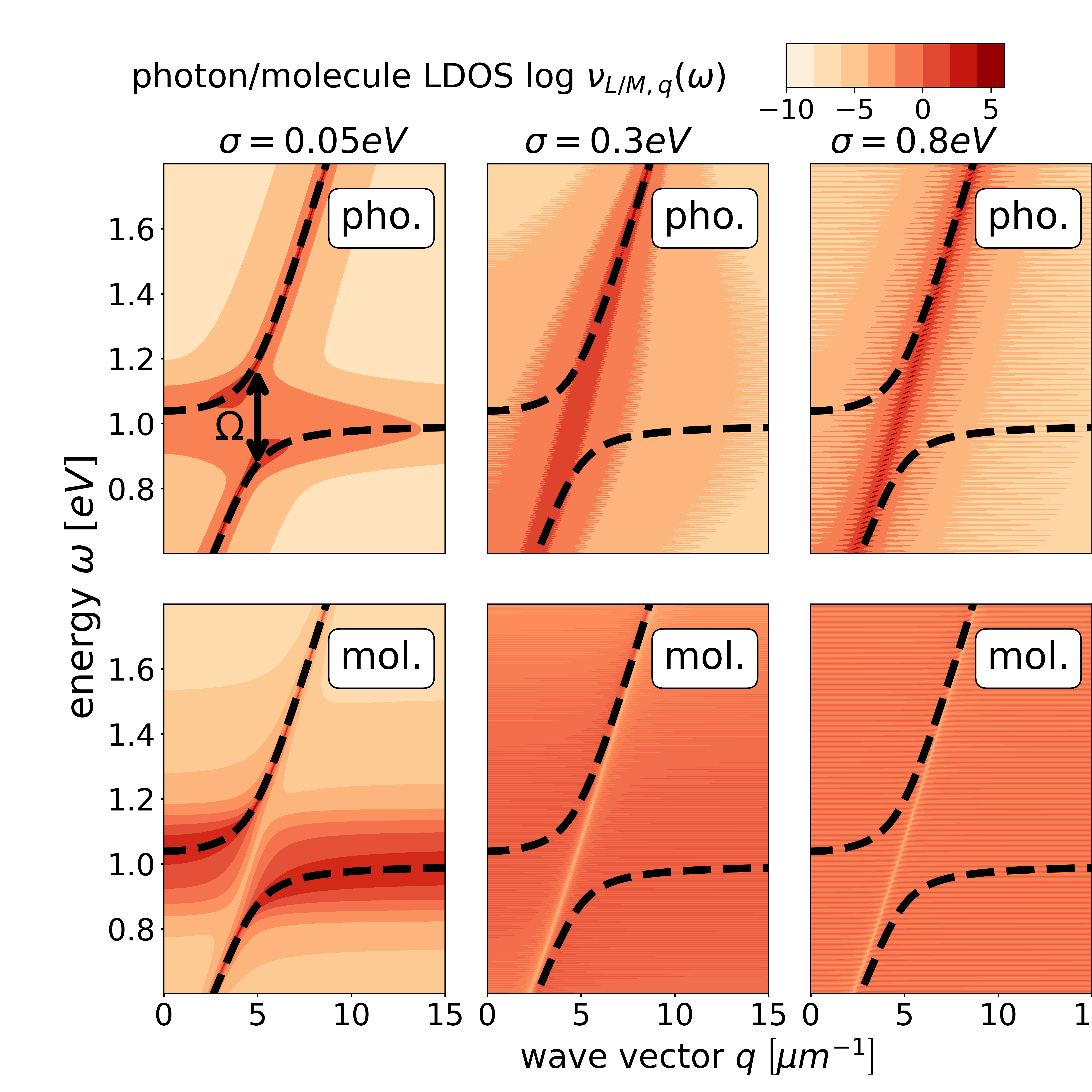}
	\caption{ Wave-vector-resolved  photon and molecule LDOS for three different disorders $\sigma$ for a Gaussian disorder distribution.   Overall parameters are $E_{C }=0.4 \,\text{eV}$,  $L = 124\,\mu \text{m}$,   $N= 5000$, $E_{\text{M} } = 1.0\,\text{eV}$, $\text{eV}$, and $g\sqrt{\rho}= 0.14\,\text{eV}$.  }
	\label{figLocalDensityOfStatesSigma_Gauss}
\end{figure}

Without disorder $\sigma=0$, all molecule energies are equal $E_j =E_{\text{M} }$ and the Hamiltonian can be easily diagonalized. Transforming the molecule operators into the wave-vector space,  $\hat B_k  = \frac{1}{\sqrt{N}}\sum_{j=1}^{N} e^{i q_k r_j} \hat B_j$, we find that the Hamiltonian is block diagonal in $k$ and reads as
\begin{eqnarray}
\hat H &=& \sum_k \hat H_k, \nonumber\\
\hat H_k  &=&  E_{\text{M} } \hat B_{k}^\dagger \hat B_k +  \omega_{ k}   \hat a_{ k }^\dagger\hat a_{ k} + g \sqrt{\rho} \left[ \hat B_{k}^\dagger    \hat  a_{ k} +\text{H.c.} \right],
\end{eqnarray}
where we have assumed a constant $g_k$ for simplicity.
For each $k$,  the two energies correspond to the lower and upper polaritons and are given as
\begin{equation}
E_{k,lo/up} = \frac{\omega_k + E_{\text{M} }}{2} \mp \frac{1}{2}\sqrt{\left( \omega_k - E_{\text{M} }\right)^2 + \Omega^2 },
\end{equation}
where we have defined the Rabi splitting of the disorder-free system $\Omega = 2 g\sqrt{\rho}$. The dispersion relations of the lower and upper polariton branches are depicted in Figs. \ref{figLocalDensityOfStates} - \ref{figLocalDensityOfStatesSigma_Gauss} by dashed lines. The corresponding eigenstates are
\begin{eqnarray}
\left|\Psi_{lo}(k) \right> &=& \left[ \cos (\theta) \hat a_{ k }^\dagger + \sin(\theta)\hat B_{ k }^\dagger \right] \left| \text{vac} \right> ,\nonumber \\
\left|\Psi_{up}(k) \right>  &=&\left[ -\sin( \theta) \hat a_{ k }^\dagger+ \cos(\theta)\hat B_{ k }^\dagger \right] \left| \text{vac} \right>,
\end{eqnarray}
where 
\begin{equation}
\theta = \frac{1}{2}\arctan \left[ \frac{g_k \sqrt{\rho} }{\omega_k - E_{\text{M} }  }\right] + \frac{\pi}{2} \theta\left(\omega_k - E_{\text{M} }\right)
\end{equation}
with the Heaviside function $\Theta(x)$.

Crucially, all molecule operators  $\hat B_k$ are coupled to the photonic operators $\hat a_k$ with the same wave vector. Thus, in contrast to single-mode models, there are no dark states in the system. However, for very large $k$, the photonic energy $\omega_k$ by far exceeds the molecule excitation energy  $ E_{\text{M} }$ such that $E_{k,lo} \rightarrow E_{\text{M} }$ and $\theta \rightarrow \pi/2$. In this limit, the photon modes and molecule modes $k$ are nearly decoupled such that the molecule mode $k$ can be considered as \textit{dark}. We follow the approach suggested in  Ref.~\cite{Ribeiro2022} and classify an eigenstate according to its photon  weight as dark or bright. Without disorder, the photon and molecule weights of the lower polariton are defined by
\begin{eqnarray}
W^{(\text{L})} (E_{k,lo}) = \sin^2(\theta), \nonumber \\
W^{(\text{M})} (E_{k,lo}) = \cos^2(\theta) .
\end{eqnarray}
Accordingly, one can define the weights for the upper polariton.  As suggested by Ref.~\cite{Ribeiro2022}, an eigenstate is classified as dark, when its photon weight is below a threshold value. In this work, we adopt the threshold value $ W^{(\text{L})}_{th} (E) = 10\%$. For disordered systems, a general definition of the photon and molecule weights is given in Eq.~\eqref{eq:def:photonMatterWeights}.

\subsection{Heisenberg equations of motion}

The analytical solution for  the system operators $\hat B_j $ and $\hat 	a_{k}$ can be obtained in Laplace space. This solution can then be used to construct arbitrary   Green's functions. The Heisenberg equations for the operators in the multimode Tavis-Cummings model read as
\begin{eqnarray}
\partial_t 	\hat B_j = - i E_j \hat B_j - i  \sum_k g_{j, k } \hat a_{ k}, \nonumber \\
\partial_t  \hat 	a_{ k}   = - i \omega_{ k} \hat a_{ k}   - i  \sum_k g_{j, k} \hat  B_j .
\end{eqnarray}
Transforming into the Laplace space defined by $\hat f(z) = \int_{0}^{\infty} dt e^{-zt} \hat f(t) $ for arbitrary   operators $\hat f(t)$, the equations of motions become
\begin{eqnarray}
z 	\hat B_j - \hat B_j^{(0)}  = - i E_j \hat B_j - i  \sum_{ k} g_{j,k } \hat a_{ k} , \label{eq:bjLaplaceSpace}\nonumber\\
z  \hat a_{ k}  -\hat a_{ k}^{(0)}   = - i \omega_{ k} \hat a_{ k }   - i  \sum_{j=1}^N g_{j,k}^* \hat  B_j  \label{eq:akLaplaceSpace},
\end{eqnarray}
where $\hat B_j^{(0)} =\hat B_j (0)$ and $\hat a_{ k}^{(0)}   =\hat a_{ k}(0)  $ denote the initial conditions of the operators at time $t=0$.
In general, this set of coupled linear equations can not be solved analytically for a large  number of photonic modes. However, we can find an exact solution in the thermodynamic limit. To see this, we first resolve Eq.~\eqref{eq:bjLaplaceSpace} and obtain
\begin{equation}
\hat B_j  =\frac{\hat B_j  ^{(0)}  }{z + i E_j}- i  \sum_{ k }  \frac{g_{j, k }\hat a_{ k}  }{z + i E_j}  .
\label{eq:formalSolutionB}
\end{equation}
Inserting this into Eq.~\eqref{eq:akLaplaceSpace} and resolving for $\hat a_{k}  $, we find
\begin{eqnarray}
\hat a_{ k}   &=& \frac{\hat a_{ k}  ^{(0)} }{z + i \omega_{ k }(z ) }  - i  \sum_{j=1}^N   \frac{g_{j, k }^* \hat B_j ^{(0)}  }{\left[z + i \omega_{ k} (z ) \right] \left(z + i E_j\right)} \nonumber \\
&-&  \sum_{k_1\neq k} \sum_{j=1}^N \frac{ g_{j, k}^*   g_{j, k_1}  \hat a_{ k_1}   }{\left[ z + i \omega_{k} (z ) \right]\left(z + i E_j\right)},
\label{eq:akExactSolution}
\end{eqnarray}
where we have defined
\begin{eqnarray}
\omega_{ k}  (z) &=& \omega_{ k}   -i \sum_{j=1}^N  \frac{ \left|  g_{j, k }\right|^2     }{z + i E_j}  \nonumber \\
&=&  \omega_{ k}   -i  \Pi(z). \label{eq:omegaRetarded}
\end{eqnarray}
where $ \Pi(z)$ will be denoted as self-energy.
The third term in Eq.~\eqref{eq:akExactSolution} represents an all-to-all coupling of the photonic modes, which cannot be solved in general. Fortunately, this term vanishes in the thermodynamic limit $N\rightarrow \infty$, as we explain in the following. To this end, we  consider the factors
\begin{eqnarray}
A_{k,k_1} &=&  \sum_{j=1}^{N} \frac{ g_{j,k }^*  g_{j, k_1}    }{z + i E_j}\nonumber \\
&=&  \sum_{j=1}^{N} \frac{g_{ k }^* g_{  k_1}      }{z + i E_j}  \varphi_{ k }^*( r_j )   \varphi_{k_1 }( r_j )\nonumber  \\
&=&  \frac{g^2}{L}\sum_{j=1}^{N} x_j + i y_j.
\end{eqnarray}
In the second equality, we have inserted the parameterization $g_{j,\kappa} = g_{ k}    \varphi_{ k }(r_j )$. In the third line, we have introduced $g$  as a typical measure for $g_{k} $. To explain the scaling of $A_{k,k_1}$, we have excluded the cavity length $L$ normalizing  the photonic mode functions $\varphi_{ k }( r )\propto 1/\sqrt{L}$. 

Because of the energetic disorder, the real and imaginary parts $x_j$ and $y_j$ are samples of random variables $X_j$ and $Y_j$, respectively. 
According to the Gaussian law of large numbers, the means and the variances of the accumulated random variables scale as
\begin{eqnarray}
\left< \sum_{j=1}^{N} X_j \right>& \propto &  \left<\sum_{j=1}^{N} Y_j \right> \propto  \delta_{k_1,k  } ,\nonumber\\
\text{Var}\; \sum_{j=1}^{N} X_j  & \propto&   \text{Var}\; \sum_{j=1}^{N} Y_j \propto  N.
\end{eqnarray}
Thus, the expectation values vanish except for $k_1= k$, while the variances scale linearly with $N$. Consequently, for $k_1\neq  k$ we find  $ A_{k,k_1} \propto  g^2\sqrt{N}/L =  g^2  \rho / \sqrt{N} $. Thus, when approaching the thermodynamic limit $N,L\rightarrow \infty$ while keeping the density $\rho$ constant, the terms $A_{k,k_1}\rightarrow 0$ with $k\neq k_1$ vanish.

Combining Eq.~\eqref{eq:formalSolutionB} with Eq.~\eqref{eq:akExactSolution}, in which we neglect the third term, we obtain the following  solution for the photonic modes and molecule excitations
\begin{eqnarray}
\hat a_{k}(z)   &=& \frac{\hat a_{k}^{ (0)} }{z + i \omega_{k} (z )  }  - i  \sum_j  \frac{g_{j,k } \hat B_j^{ (0)} }{\left[ z + i \omega_{k}(z ) \right] \left(z + i E_j\right)}  \nonumber ,  \\
\hat B_j(z)   &=& \frac{\hat B_j^{ (0)}  }{z + i E_j} 
-i\sum_{k}  \frac{g_{j,k } \hat a_{k}^{ (0)} }{\left(z + i E_j\right) \left[ z + i \omega_{k}(z ) \right] } \nonumber  \\ 
&-&  \sum_{k}   \sum_{j_1}  \frac{ g_{j,k }  g_{j_1,k }^* \hat B_{j_1}^{ (0)}  }{\left(z + i E_j\right) \left[ z + i \omega_{k} (z ) \right] \left(z + i E_{j_1}\right)} , 
\label{eq:operatorsLaplaceSpace:app}
\end{eqnarray}
respectively.

\subsection{Green's function}

Based on Eq.~\eqref{eq:operatorsLaplaceSpace:app}, one  can directly obtain the retarded Green's functions defined by
\begin{eqnarray}
G_{k,k^\prime }^{(\text{L})}(z) &\equiv&- i \left<\left[ \hat a_{k}(z) , \hat a_{k^\prime}^{(0)\dagger}\right] \right>  ,\nonumber \\
G_{j,j'}^{(\text{M})}(z) &\equiv &-i \left< \left[ \hat B_j(z) , \hat B_{j'}^{(0)\dagger}\right] \right>,
\label{eq:def:GreensFkt}
\end{eqnarray}
for photonic operators and molecule operators, respectively. As we consider bosonic operators in a non-interacting system, the expectation value in Eq.~\eqref{eq:def:GreensFkt} does not depend on the initial  condition, which we do not specify for this reason.  Explicitly, the photonic and molecule Green's functions read as
\begin{eqnarray}
G_{k,k^\prime }^{(\text{L})}(z)   &=& \delta _{k,k^\prime } \frac{-i }{z + i \omega_{k} (z )  }   \nonumber ,  \\
G_{j,j'}^{(\text{M})}(z)   &=& \frac{-i   }{z + i E_j}  \delta_{j,j^\prime}\nonumber  \\ 
&+& i \sum_{k}    \frac{ g_{j,k }  g_{j^\prime,k }^*   }{\left(z + i E_j\right) \left[ z + i \omega_{k} (z ) \right] \left(z + i E_{j^\prime}\right)} .
\label{eq:greensFunction}
\end{eqnarray}
Similarly, mixed light-matter Green's functions can be constructed. 
Using the photonic mode functions, we can express the photon Green's function in  position space as
\begin{eqnarray}
G_{j ,j^\prime }^{(\text{L})}(z)   &=& i \sum_{  k}\frac{ \varphi_{k}( r_j )  \varphi_{ k}^{*} (  r_{j'} ) }{z + i \omega_{ k } (z )  }   .
\label{eq:photonGreensFunction}
\end{eqnarray}

\subsection{Disorder average}

We define the disorder-averaged Green's function as
\begin{eqnarray}
&&\overline G_{a,a^\prime }^{(X)} \left( z\right) 
=  \lim_{N \rightarrow\infty  }  \int   \left[ \prod_{i=1}^{N}   dE_{i} P(E_i) \right]  G_{a,a^\prime}^{(X)} (z),
\label{eq:disorderAveragedGreensFct}
\end{eqnarray}
where $X\in \left\lbrace \text{L},\text{M}\right\rbrace$, and  $a,a^\prime$ can label position $j$ or wave vector $k$. 

\subsubsection{Self-energy}

We observe that the photonic Green's function in Eq.~\eqref{eq:greensFunction} depends on the disorder  via the self energy $ \Pi(z)$ in the renormalized frequencies $\omega_{k} (z )$ in Eq.~\eqref{eq:omegaRetarded}. Using $\left|  g_{k,j }\right|^2  =  g_{k }^2/L  $, the self-energy becomes
\begin{eqnarray}
\Pi(z) &=& \lim_{N \rightarrow\infty  }	 \sum_{j=1}^N  \frac{ \left|  g_{k,j }\right|^2     }{z + i E_j}   \nonumber \\
&\rightarrow&  N \int  dE P(E)  \left [\frac{1}{L}  \frac{ \left|  g_{  k}      \right|^2     }{z + i E}  \right ]  \nonumber \\
&=& i \left|  g_{ k}\right|^2  \rho  \Gamma(z).
\end{eqnarray}
As the summation in the first line does not depend on the position, the sum over $N$ can be interpreted as an integral over the energy weighted by the disorder distribution $P(E)$ for $N\rightarrow \infty$. Finally, we have introduced the disorder-averaged Green's function of the uncoupled molecules 
\begin{equation}
\Gamma(z) = - i\int dE \frac{P(E)}{ z + i E  } .
\label{eq:freeQuantumEmitterGreenFkt}
\end{equation}
The key point in this derivation is that the self-energy term is self-averaging in the thermodynamic limit and thus does not require the  external averaging operations defined in Eq.~\eqref{eq:disorderAveragedGreensFct}.

\subsubsection{Disorder-averaged Green's function}

Given that the self energy is self averaging, it is now straight forward to perform the disorder average of the Green's function in Eq.~\eqref{eq:greensFunction}. Apart from $\omega_k(z)$, the photon Green's function does not depend on the molecule energies such that the disorder-averaged Green's functions coincides with the expressions in Eqs.~\eqref{eq:greensFunction} and \eqref{eq:photonGreensFunction} in position and wave vector space, respectively.

The disorder average of the matter function has the effect that the terms $1/(z+i E_j)$ are replaced by $i\Gamma(z)$ in Eq.~\eqref{eq:freeQuantumEmitterGreenFkt}, i.e., 
\begin{eqnarray}
G_{j,j^\prime}^{(\text{M})} \left(z\right)  &=&   \left[ \Gamma(z) \delta_{j,j^\prime}  \phantom{ \frac{ \varphi_{ k}( r )  \varphi_{ k}^{*} ( r^\prime )  }   { z + i \omega_{  k} (z )  }} \right.  \nonumber \\
&-&\left.  i \Gamma^2(z) \sum_{  k}   \left| g_{ k}\right|^2 \frac{ \varphi_{ k}( r_{j} )  \varphi_{ k}^{*} ( r_{j^\prime} )  }   { z + i \omega_{ k} (z )  } \right],
\label{eq:greenFkt_matterPosition}
\end{eqnarray}
which in momentum space  becomes
\begin{equation}
G_{k,k ^\prime }^{(\text{M})}( z )  =   \delta _{k,k ^\prime } \left[   \Gamma(z)  -i  \Gamma^2(z)     \frac{   \left| g_{  k}\right|^2 \rho }   { z + i \omega_{ k} (z )  } \right].
\label{eq:greenFkt_matterMomentum}
\end{equation}
We emphasize that these Green's functions are exact in the  thermodynamic limit $N\rightarrow \infty$ because of the self-averaging property of the self energy. Formally, the self average is equivalent to the celebrated coherent potential approximation (CPA). However, we emphasize that the CPA is exact for arbitrary disorder distributions for the model considered here. This is in contrast to nearest-neighbor and other short-range hopping models, where the CPA is only exact for the Lorentzian disorder model~\cite{Chenu2017}.

Noteworthy, the Green's function in wave vector space $G_{k,k ^\prime }^{(\text{L})}( z ) $ in Eq.~\eqref{eq:greensFunction} and $G_{k,k ^\prime }^{(\text{M})}( z ) $ in Eq.~\eqref{eq:greenFkt_matterMomentum} are identical to the Green's function of the single-mode Tavis-Cummings model when replacing the photonic dispersion relation $\omega_k$ by the energy of the single cavity mode $\omega_C$. The single-mode model with Lorentzian disorder has been  investigated in detail in Ref.~\cite{Engelhardt2022}. This shows that spectroscopic quantities, which can be wave-vector-resolved measured, can be correctly calculated using  single-mode models. In contrast, the Greens's functions in positions space given in Eq.~\eqref{eq:photonGreensFunction} and \eqref{eq:greenFkt_matterPosition} involve a summation of the wave vector. Therefore, we  conclude that transport quantities and the  coherence length cannot be accurately investigated in  single-mode models. Thereby, the wave-vector summation in Eq.~\eqref{eq:photonGreensFunction} and \eqref{eq:greenFkt_matterPosition} ensures that the speed-of light is maintained as an upper bound in the dynamics.

\subsection{Photon and molecule local density of states}

In terms of the eigenstates $\left|\alpha\right>$ and energies $\epsilon_\alpha$ of the Hamiltonian in Eq.~\eqref{eq:Hamiltonian}, we define the wave-vector-resolved  photon and molecule local density of states (LDOS) via
\begin{eqnarray}
\nu_{\text{L},k}(\omega)  &=&  \frac{1}{2\delta} \sum_{\epsilon_{\alpha} \in \left[\omega-\delta,\omega+\delta \right] } \left< \alpha \right| \hat a_k^\dagger \hat a_k \left|\alpha \right> \nonumber ,\\
\nu_{\text{M},k}(\omega)  &=&  \frac{1}{2\delta}\sum_{\epsilon_{\alpha} \in \left[\omega-\delta,\omega+\delta \right] } \left< \alpha \right| \hat B_k^\dagger \hat B_k \left|\alpha \right> 
\label{eq:def:LDOS}
\end{eqnarray}
with an infinitesimal $\delta>0$. We note that the photon and molecule LDOS can be measured by angular-resolved spectroscopy. The diagonal elements of the Green's function in Eq.~\eqref{eq:def:GreensFkt} can be formally written as
\begin{eqnarray}
G_{k,k }^{(\text{L})}(z) &\equiv& -i \sum_\alpha   \frac{\Psi_{\alpha}^{(\text{L})}(q_k)  \Psi_{\alpha}^{(\text{L})*}(q_k) }{z+ i \epsilon_\alpha}  ,\nonumber \\
G_{k,k}^{(\text{M})}(z) &\equiv &-i\sum_\alpha   \frac{\Psi_{\alpha}^{(\text{M})}(q_k)   \Psi_{\alpha}^{(\text{M})*}(q_k)  }{z+ i \epsilon_\alpha},
\label{eq:def:GreensFktFormal}
\end{eqnarray}
where $ \Psi_{\alpha}^{(\text{L})}(q_k)   = \left< \alpha \right|  \hat a_k^\dagger  \left| \text{vac}\right> $ and $\Psi_{\alpha}^{(\text{M})}(q_k)  = \left< \alpha \ \right|  \hat B_k^\dagger  \left|\text{vac}\right> $ are the eigenstates in wave vector representation. Using the Dirac identity $\lim_{\delta\downarrow 0}1/(x+ i\delta) = \mathcal P\, 1/x -i\pi \delta(x)$, it is now straightforward to show that 
\begin{equation}
\nu_k^{(X)}(\omega) =- \lim_{\delta\downarrow 0} \frac{1}{\pi}\,\text{Im}\, G_{k,k }^{(X)}(-i\omega+\delta)
\end{equation}
for $X\in \left\lbrace \text{L},\text{M}\right\rbrace $.  For later purpose, we also define the total density of states
\begin{eqnarray}
\nu(\omega) &\equiv &  \frac{N_{\omega,\delta}}{2\delta}  \nonumber \\
& = &  \sum_{k} \left[ \nu_{\text{L},k}(\omega) + \nu_{\text{M},k}(\omega)\right],
\label{eq:def:totalDOS}
\end{eqnarray}
where  $N_{\omega,\delta}$ is the number of eigenstates in the energy interval $\left[\omega-\delta,\omega+\delta \right]  $ having an infinitesimal width $2\delta$. The equality in the second line follows from the fact that $\sum_{k}\left[ \hat a_k^\dagger \hat a_k + \hat B_k^\dagger \hat B_k   \right] =\mathbbm 1$ in the single-excitation manifold. 

In Fig.~\ref{figLocalDensityOfStates}, we analyze the photon and molecule LDOSs for three confinement energies $E_{C } = 0.4\,\text{eV}$, $E_{C } = 1.0\,\text{eV}$ and $E_{C } = 1.3\,\text{eV}$. For simplicity we assume a wave-vector-independent light-matter interaction $g_k=g$. All three photon and molecule LDOSs look qualitatively similar. Yet, we find that the photon LDOS for $E_{C } = 1.3\,\text{eV}$ has a significant smaller contribution for energies close to $E_{\text{M} }$ than the ones for $E_{C } = 0.4\,\text{eV}$ and $E_{C } = 1.0\,\text{eV}$. In contrast, the matter LDOS for $E_{C } = 1.3\,\text{eV}$ has a significant smaller contribution close to the photon dispersion  $\omega_{k }$ than the ones for $E_{C } = 0.4\,\text{eV}$ and $E_{C } = 1.0\,\text{eV}$.

In Fig.~\ref{figLocalDensityOfStatesLMintLarge}, we analyze the LDOSs for the same parameters as in Fig.~\ref{figLocalDensityOfStates}, but for a  larger light-matter coupling. Accordingly, we see that now the Rabi splitting $\Omega = 2 \sqrt{\rho} g$ is significantly larger. Consequently, the transparency effect in the matter LDOS within the gap is better visible.

Next, we investigate the influence of  disorder on the photon and molecule LDOSs, which is depicted in Fig.~\ref{figLocalDensityOfStatesSigma_Gauss} for three different $\sigma $. For $\sigma = 0.05 \,\text{eV}$, we observe two separate polariton bands, that merge for $\sigma  = 0.3  \,\text{eV}$. When further increasing to $\sigma= 0.8  \,\text{eV}$, the photon LDOS increasingly concentrates around the  photon dispersion relation $\omega_k$: for increasing $\sigma$, the molecular excitation energies are distributed over a larger energy range,  which leads to an  increasing decoupling of photonic and molecular systems, analog to the behavior in single-mode models~\cite{Engelhardt2022}.

\subsection{Photon and molecule weights}

\begin{figure}[t]
	\includegraphics[width=\linewidth]{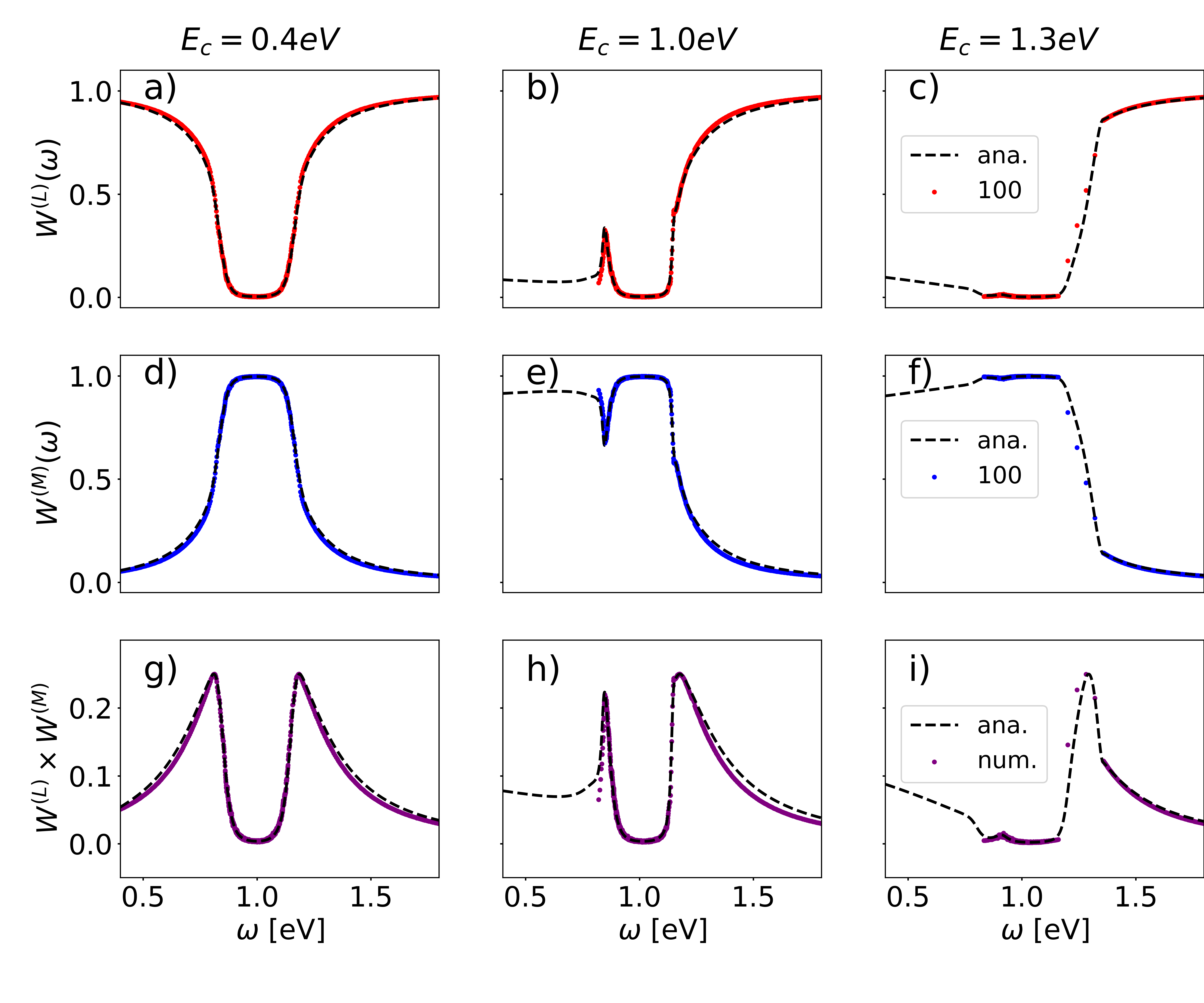}
	\caption{ Investigation of the photonic weight $W^{(\text{L})}(\omega ) $ [panels (a), (b), and (c)]  and matter weight $W^{(\text{M})}(\omega) $  [panels (d), (e), and (f)]  defined in Eq.~\eqref{eq:disorederAveragedWeights}.  Panels (g), (h), and (i) depict the product $W^{(\text{L})}(\omega )\cdot W^{(\text{M})}(\omega ) $ as a measure for the light-matter mixing. Overall parameters are the same as in Fig.~\ref{figLocalDensityOfStates}.}
	\label{figLightMatterMixing}
\end{figure}

The  photon and  molecule weights of a polariton determine its dynamical properties. In terms of the eigenstates $\left| \alpha\right>$ of the multimode Tavis-Cummings Hamiltonian, these quantities are defined as
\begin{eqnarray}
W^{(X)}(\epsilon_\alpha)  &\equiv&   \left< \alpha \right| \sum_k \hat a_{k}^\dagger \hat a_k   \left| \alpha \right>,
\label{eq:def:photonMatterWeights}
\end{eqnarray}
where $X\in \left\lbrace \text{L}, \text{M}\right\rbrace$.
In the numerical calculations, we evaluate the averaged quantities
\begin{equation}
W^{(X)}(\omega)  = \frac{1}{N_{\omega,\delta}} \sum_{\epsilon_{\alpha} \in \left[\omega-\delta,\omega+\delta \right] } W^{(X)}(\epsilon_\alpha) ,
\label{eq:disorederAveragedWeights}
\end{equation}
where  $N_{\omega,\delta}$ is the number of eigenstates in the energy interval $\left[\omega-\delta,\omega+\delta \right]  $ of width $2\delta\ll \omega$. Note that due to their definition, the photon and matter weights sum up to one, i.e., $W^{(\text{L})}(\omega) + W^{(\text{M})}(\omega)= 1$.  Using the definitions of the wave-vector-resolved LDOS in Eq.~\eqref{eq:def:LDOS} and the total density of states in Eq.~\eqref{eq:def:totalDOS}, we find
\begin{eqnarray}
W^{(X)}(\omega )  &=& \sum_k \frac{\nu_k^{(X)}(\omega ) }{\nu(\omega ) }.
\label{eq:rel:weightInTermsOfLDos}
\end{eqnarray}
Thus, the photon (molecule) weight is the  photon (molecule) LDOS integrated over the wave vector $k$ and normalized by the total density of states.  This expression can be numerically integrated using the analytical expression of the Greens's functions in Eq.~\eqref{eq:greensFunction} and \eqref{eq:greenFkt_matterMomentum}.

In Fig.~\ref{figLightMatterMixing} (a) - (c) we depict the  photon  weight  as a function of energy for three different confinement energies $E_{C }$. These plots serve also as a benchmark calculation of our analytical solution in Eqs.~\eqref{eq:greensFunction} and \eqref{eq:greenFkt_matterMomentum}. Thereby, the weights have been numerically evaluated using Eq.~\eqref{eq:disorederAveragedWeights}, where the eigenstates $\left| \alpha\right>$ have been obtained by numerical diagonalization of the Hamiltonian. In panels (a)- (c) we observe a significant dip in the photon weight for energies around $\omega \approx E_{\text{M} }$. As the photon weight is less than $W^{(\text{L})}< 10\% $, these eigenstates are classified as dark states according to the explanations in Sec.~\ref{sec:noDisorder}. In panel (b) and (c) we find that the photon weight is larger than $W^{(\text{L})}= 50\% $ only for energies above $ E_{\text{M} }$, as the photonic dispersion relation is bounded by $E_{C }$ from below.
The panels   Fig.~\ref{figLightMatterMixing} (d) - (f) depict the corresponding matter weights. 
Moreover, we depict the light-matter mixing, which we define as the product $W^{(\text{L})}(\omega) \cdot W^{(\text{M})}(\omega)$. The peaks of  this quantity clearly resembles the lower and upper polariton  bands.

\subsection{Asymptotic behavior in position space}

\begin{figure}[t]
	\includegraphics[width=\linewidth]{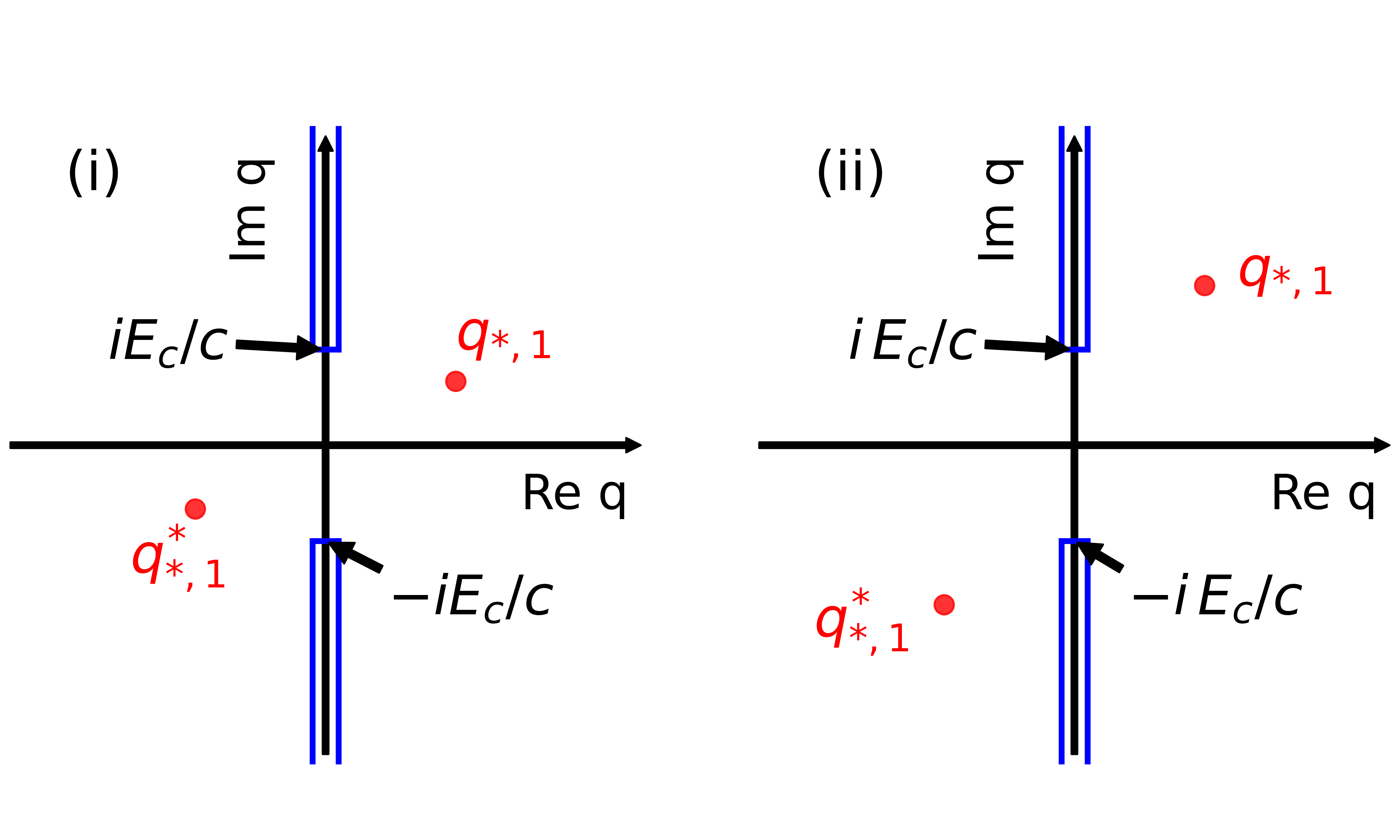}
	\caption{Sketch of the nonanalyticities of the Green's function in the complex $q$ plain, that determine the asymptotic behavior of the Fourier transformation in Eq.~\eqref{eq:photonGreensFunctionContLim}. The branch cut of the integrand $G_q$ are depicted by blue contours. The poles of $G_q$ are marked by red points. }
	\label{figRootsAnalysis}
\end{figure}

Next, we are interested in the asymptotic behavior of the Green's function in position space, i.e., we  investigate $ G_{j,j^\prime}^{(X)} \left(-i\omega^+\right) $ for $X\in \left\lbrace \text{L},\text{M} \right\rbrace$ and large $\left| r_j -r_{j^\prime } \right| $, where $\omega^+= \omega+i\delta$ with infinitesimal $\delta>0$. From Eqs.~\eqref{eq:greensFunction} and \eqref{eq:photonGreensFunction} we see that the molecule Green's function is proportional to the photon Green's function when $g_{k}$ changes only slowly with $k$. We thus continue to investigate the latter.  In the continuum limit $L\rightarrow\infty$, the photon Green's function in position space [c.f., Eq.~\eqref{eq:photonGreensFunction}]  reads as 
\begin{eqnarray}
G_{j ,j^\prime }^{(\text{L})}\left(-i\omega^+\right)   &=&  \int_{-\infty}^{\infty }G_q(-i\omega )  e^{iq \left( r_j -r_{j^\prime} \right) }  dq  , \nonumber \\
G_q(-i\omega )   &=& \frac{ -i }{ -i\omega + i \omega_{ qL/2\pi } +  i g_{  qL/2\pi}^2  \rho  \Gamma(-i\omega) },\nonumber \\
\label{eq:photonGreensFunctionContLim}
\end{eqnarray}
where we can replace $\omega^+ \rightarrow \omega$ as the poles of the Green's function are not  located on the real axes because of the complex-valued $\Gamma(z)$. 

We can analyze the asymptotic behavior of the Green's function using a theorem of  functional analysis~\cite{Reed1975}:

\begin{quote}
	Let $f$ be in $L^2\left( \mathbbm{R}^n\right)$ (space of square-integrable functions). Then $e^{b\left|x \right|}f \in L^2\left(  \mathbbm{R}^n\right)$ for all $b< a$ if and only if its Fourier transformation $\tilde f$ has an analytic  continuation to the set $\left\lbrace \zeta \mid \left| \zeta \right|<a \right\rbrace $ with the property that for each $\eta \in  \mathbbm{R}^n$ with $\left| \eta \right| < a$, $\tilde f (\cdot + i \eta)\in L^2\left(\mathbbm R^n\right)$ and for any $b<a$: $\sup_{\left|\eta\right|\leq b}  \left|\left|f(\cdot + i\eta )\right|\right|_2<\infty$.
\end{quote}
Applied to the Fourier transformation in Eq.~\eqref{eq:photonGreensFunctionContLim}, this  theorem states that the asymptotic behavior of the Green's function is determined by the poles and branch cuts of the integrand (i.e., the Green's function in wave vector space), where $q$ is now interpreted as a complex variable. 

For simplicity, we consider the case $g_k =g$. In this case the integrand $G_q$ is nonanalytic at
\begin{eqnarray}
q_{*,1}  &=& \frac 1 c \sqrt{ \left[ \omega -  g^2 \rho \Gamma(-i\omega) \right]^2 -E_{C }^2 } \nonumber, \\
q_{*,2} &=&  i \frac{ E_{C }}{c} ,
\label{eq:greensFktNonAnalyticities}
\end{eqnarray}
as well as the complex conjugate values. Thereby, $q_{*,1}$ appears because of the pole of the Green's function $G_q$,   $q_{*,2}$ appears because of the branch cut induced by the photonic dispersion relation in Eq.~\eqref{eq:photonDispRelation}.  According to the above theorem, the asymptotic behavior is mainly determined by the nonanalyticity whose imaginary part is closer to zero.

In Fig.~\ref{figRootsAnalysis}, we sketch two  cases: (i)  $\left|\text{Im }\, q_{*,1}  \right|  < \left|\text{Im }\,q_{*,2}  \right|  $, and (ii)   $\left|\text{Im }\, q_{*,1}  \right|  > \left|\text{Im }\,q_{*,2}  \right|  $. In case (i),  we  find that the asymptotic decay is determined by the pole of the Green's function. In this case, the coherence length depends on the light-matter interaction and other system parameters.  In case (ii), the branch cut and, consequently, the ratio $E_{C }/c$ determines the asymptotic behavior. In short, the asymptotic behavior for large  $r= \left|r_j -r_{j^\prime } \right|$ can be written as
\begin{equation}
\left| G_{j ,j^\prime }^{(\text{L})}\left(-i\omega^+\right)\right| \rightarrow a_1 e^{-\frac {r}{2\zeta_{\text{coh},1}  } } + a_2 e^{-\frac {r}{2\zeta_{\text{coh},2}  } },
\label{eq:greensFktAsymptoticBehavior}
\end{equation}
where we have defined the coherence lengths
\begin{equation}
\zeta_{\text{coh},i} = \frac{1}{ 2 \text{Im } q_{*,i}  }
\label{eq:coherenceLength}
\end{equation}
for $i=1,2$.
In the numerical calculations we find that  the asymptotic behavior is accurately determined by $q_{*,1}$, see Sec.~\ref{eq:disorderAveragedGreensFctNum}.  For this reason,  we assess that the coefficients fulfill $a_1\gg a_2$, and we define the system's coherence length as  $\zeta_{\text{coh}} \equiv \zeta_{\text{coh},1}$.

In the Letter, the coherence length for different values of the confinement energy $E_{C }$ is shown in Figs.~1(g)-(i). Comparing with the photon weight, we observe a clear correlation between both quantities, that will be explained in the next section. Consequently, the coherence length is very short in the energy range close to $\omega \approx E_{\text{M} }$, that exhibits dark states. From this observation we conclude that dark states have an overall destructive impact on transport properties.

It is also instructive to investigate the coherence length for a small light-matter interaction. Taylor expansion of Eq.~\eqref{eq:greensFktNonAnalyticities} up to first order in $g^2\rho$ results in
\begin{eqnarray}
\zeta_{\text{coh}}^{-1} &   =  &  \frac 1 c \text{Im} \left[ \sqrt{\omega^2 - E_{C}^2} -  \frac{\omega }{ \sqrt{\omega^2 - E_{C}^2} } g^2 \rho  \Gamma(-i\omega) \right] \nonumber \\
&     & +  \mathcal O \left[ \left(   g^2 \rho\right)^2    \right],
\label{eq:res:localiationLengthExpansion}
\end{eqnarray}
which exhibits a different behavior depending on the energy $\omega$. For $\omega> E_{C }$, the first term in Eq.~\eqref{eq:res:localiationLengthExpansion} is real valued and the coherence length is essentially determined by the imaginary part of $\Gamma(-i\omega)$. This analysis thus reveals why the coherence length decreases with $\propto (g^2\rho)^{-1}$ in Fig. 3(a) and (c) in the Letter for $\omega>E_{C }$. As the imaginary part of $\Gamma(-i\omega)$ is given by the molecular disorder distribution $P(E)$, the coherence length is proportional to the number of molecules having excitation energy $E_j =\omega$.  Noteworthy, for $E_{C } =0 $, the inverse coherence length 
\begin{eqnarray}
\zeta_{\text{coh}}^{-1} &   \propto  &  g^2 \rho P(\omega)
\end{eqnarray}
is proportional to Beer's absorption length, which is consistent with the unraveling of the Green's function in terms of scattering processes below in Sec.~\ref{sec:interpretation}.

For    $\omega< E_{C }$, the first term becomes imaginary and gives a constant contribution to $\zeta_{\text{coh}}^{-1} $. Now, the  real part of $\Gamma(-i\omega)$ determines the dependence on $g \sqrt{\rho}$.  Interestingly when the energy  $\omega$ approaches $E_{C }$ from above, the coherence length increases. Formally for $\omega=E_{C }$, the coherence length diverges, yet, we note that  the Taylor expansion is  invalid in this case. Thus, from Eq.~\eqref{eq:res:localiationLengthExpansion} we cannot conclude that the coherence length is non-analytic at $\omega=E_{C }$.

\subsection{Interpretation}

\label{sec:interpretation}

Interestingly, the coherence length is a function of the Rabi frequency $\Omega = 2g \sqrt{\rho}$.
Analysis of the coherence length shows that it scales with $\zeta_{\text{coh}}\propto \Omega^{-2} $ for small $\Omega$, as can be seen in Figs. 3(a) and (c) in the Letter. 
In this section, we interpret this behavior in terms of photon scattering.

Expanding the photon Green's function in position space  in orders of $g$, we obtain 
\begin{eqnarray}
G_{j,j^\prime}^{(\text{L})} (z)    &=&	\sum_k G_{j,j^\prime }^{(\text{L})} (z,k) ,\nonumber    \\
G_{j,j^\prime}^{(\text{L})} (z,k)    &=&  G_{j,j^\prime}^{(0)}(k) +g^2 \sum_{j_1=1}^{N} G_{j,j_1}^{(0)}(k)  \Gamma_{j_1}G_{j_1,j^\prime }^{(0)}  (k) \nonumber    \\
& +& g^4\sum_{j_1, j_2=1}^{N} G_{j,j_1}^{(0)}(k)  \Gamma_{i_1}G_{j_1,j_2 }^{(0)}  (k) \Gamma_{i_2}G_{j_2,j^\prime  }^{(0)} (k)  \nonumber \\
& +&   \mathcal O(g^6),
\label{eq:greensFktUnraveling}
\end{eqnarray}
where
\begin{eqnarray}
G_{i,j}^{(0)} (k)   &=& G_{i,j}^{(0)} (z,k)  = - i \frac{ \varphi_{ k }( r_i )   \varphi_{ k }^{*}( r_j )}{ z + i \omega_{  k } } ,\nonumber \\
\Gamma_j &=&   \Gamma_j(z) = - i \frac{   1    }{z + i E_j}  
\end{eqnarray}
denote the free Green's functions of photon mode $k$ and  molecule $j$, respectively.
Thus, the full Green's function is a sum of $k$-dependent Green's functions, that can be unraveled as a series of scattering processes. The first term ($\propto g^0$), describes the propagation of a photon without scattering events. The second term  describes a single scattering at molecule $j_1$. Both absorption and emission contribute one factor $g$.  Noteworthy, $\text{Im}\, \Gamma_j(-i\omega ) $ is proportional to the linear absorption of a two-level systems in dipole approximation.  As  there are $N$ molecules in the cavity, this terms scales overall with $\propto g^2 N$. Accordingly, the third term can be interpreted  as two scattering events and scales with $\propto g^4 N^2$.  Each term in Eq.~\eqref{eq:greensFktUnraveling} can be  interpreted as a distinct path from $r_j$ to $r_{j^\prime}$.

For increasing $g^2 N$, more higher-order paths become relevant in the  expansion in Eq.~\eqref{eq:greensFktUnraveling}. Due to the random phase factors of $\varphi_{ k}(r_j ) $, this leads to a destructive interference between different paths, which suppresses the probability that a photon can travel from $r_j$ to $r_{j^\prime}$. This is reflected  by a shorter coherence length of the Green's function. 

The disorder distribution of the molecular excitation energies $P(E)$ thereby determines the number of molecules that can take part in this scattering process. Only molecules whose excitation energy $E_j$ exactly matches the energy of the eigenstate  $\omega (=iz)$ can resonantly scatter  photons, which mediate the formation of the eigenstates. Thereby, the more a photon is scattered at molecules the smaller is the photon weight, which explains the close correspondence of photon weight and coherence length in Fig.~1 in the Letter. This analysis motivates to identify the coherence length as the absorption length known in Beer's absorption law, which opens an alternative way for the experimental verification of the predictions in our work.

\section{Numerical calculations}

\label{sec:numericalCalculations}

\subsection{Diagonalization}

\label{sec:diagonalization}

Before describing the details of the numerical calculations, we  list here the overall parameters and procedures that are used unless stated differently. 
In the numerical calculations,  we use an open boundary condition instead of a periodic one, as the corresponding mode functions $\varphi_k(r) = \sin ( q_k r)/\sqrt{L/2}  $ with $q_k = \pi k/ L$ are real valued. This guarantees that all eigenstates are real, simplifying the disorder average. 
This is in contrast to the periodic boundary condition considered in the analytical calculations in Sec.~\ref{sec:GreensFct}. Yet, away from the boundaries, the physical properties will be the same under both boundary conditions. 

We consider a  system with $N=5000$ molecules and a cavity of length $L=124\,\mu\text{m}$. Molecule $j\in\left\lbrace1,\dots,N\right\rbrace$ is located at position $r_j = N j /L$ . For reference, we define the resonance wavelength $\lambda_0$ such that $ 2\pi c /\lambda_0 = E_{\text{M} } /\hbar$. For $E_{\text{M} } = 1\,\text{eV}$,  we find that $\lambda_0 =1.24 \,\mu\text{m} $. Thus, expressed  in units of $\lambda_0$, the  cavity length is $L =100 \lambda_0$, and the particle density is $\rho = N/L=50/\lambda_0$. We quantize the photonic field with $N= 5000$ modes, such that the cut-off energy is $\omega_{\text{cut-off}} =\omega_{k=5000} = \sqrt{c^2 ( \pi 50/L)^2+ E_{C }^2} \approx 50 \cdot E_{\text{M} }$. In total, we average over $M=50$ sample Hamiltonians.

In the diagonalization, the photon and molecule subsystems are represented in the wave vector basis and position basis, respectively. The construction of the Green's function requires  a transformation between these representations. We denote the elements of the photon subsystem  of an eigenstates $\alpha$ with energy $\epsilon_\alpha$ by $\Psi^{(\text{L})}_\alpha(q_k)$, and the  molecule subsystem elements by $\Psi^{(\text{M})}_\alpha(r_j)$. Using the photonic mode functions $\varphi_k(r)$, we can transform the photon subsystem into the position representation via
\begin{equation}
\Psi_\alpha^{(\text{L})}(r_j) = \sqrt{\frac{L}{2N^2}}\sum_{k=1}^{N}  \varphi_k(r_j) \Psi_\alpha^{(\text{L})}(q_k).
\end{equation}
The additional factor $\sqrt{L/2N}$ is required due to the discretization of the photon field in position space. It ensures that the photon weight remains unchanged in the transformation: $W^{(\text{L})}(\epsilon_a) =\sum_j \left| \Psi_\alpha^{(\text{L})}(r_j) \right|^2  = \sum_k \left| \Psi_\alpha^{(\text{L})}(q_k) \right|^2  $. Accordingly, we can transform the matter subsystem into the wave vector representation:
\begin{equation}
\Psi_\alpha^{(\text{M})}(q_k) = \sqrt{\frac{L}{2N^2}}\sum_{j=1}^{N}  \varphi_k(r_j) \Psi_\alpha^{(\text{M})}(q_k).
\end{equation}

\subsection{Disorder-averaged Green's function}

\label{eq:disorderAveragedGreensFctNum}

\begin{figure}[t]
	\includegraphics[width=\linewidth]{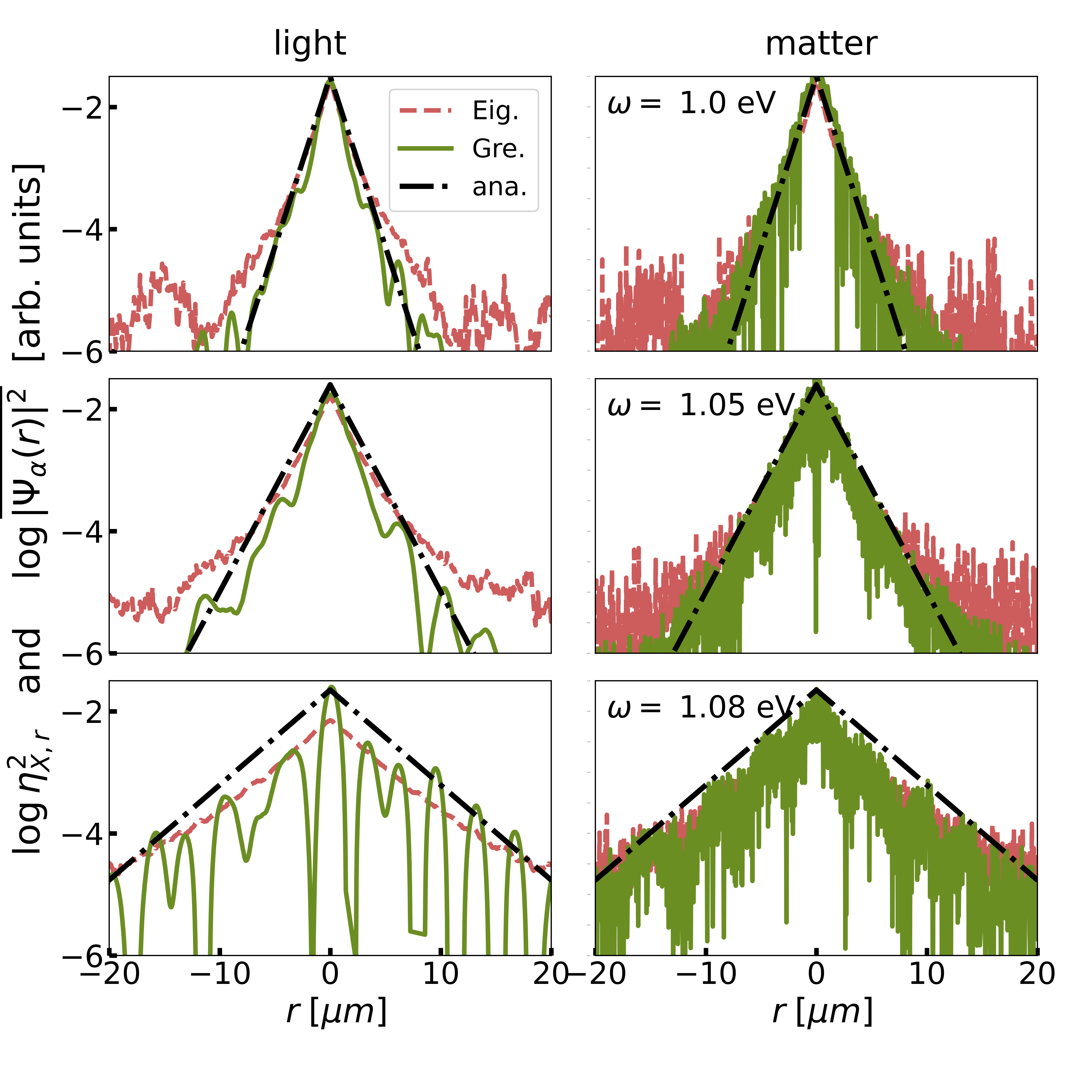}
	\caption{Comparison of $ \left[ \eta_{X,r}(\omega)\right]^2 $ [defined in Eq.~\eqref{eq:greenFctImaginaryPartSeparation} via the imaginary part of the Green's function] and the disorder-averaged eigenstate in Eq.~\eqref{eq:disordereAveragedProbabilities}. The general diagonalization procedure is described in Sec.~\ref{sec:diagonalization}. The Green's function and eigenstates are averaged over an interval of width $\delta = 0.01$. The dash-dotted black lines depict the analytical predicted exponential decay with the coherence length in Eq.~\eqref{eq:coherenceLength} for $i=1$. Parameters are the same as in Fig.~\ref{figLocalDensityOfStates} and $E_C=0.4\,\text{eV}$. }
	\label{figComparisonGreensFctWaveFct}
\end{figure}

In terms of eigenstates, the disorder-averaged  Green's function in position space can be  written as
\begin{equation}
\overline G_{j,j^\prime}^{(X)} (z) = \frac{-i}{M}\sum_{l=1}^{M} \sum_{\alpha }  \frac{ \Psi_\alpha^{(X)(l)} (r_j)   \Psi_\alpha^{(X)(l)*}(r_{j^\prime } ) }{z+i\epsilon_\alpha^{(l)} },
\end{equation}
where $\epsilon_\alpha^{(l)}$ and $ \Psi_\alpha^{(X)(l)} (r_j)$ are the energies and the eigenstates of the $l$-th sample Hamiltonian.
In the following, we  use that the Hamiltonian in Eq.~\eqref{eq:Hamiltonian} is time-reversal invariant, implying that there exist a gauge for which the eigenstates are real valued. For this reason,  it is possible to express the imaginary part of the Green's function as
\begin{eqnarray}
\eta_{j,j^\prime }^{(X)} (\omega)&\equiv& \lim_{\delta\downarrow 0}\frac{-1}{\pi}	\text{Im}\, \overline G_{j,j^\prime}^{(X)}  (-i\omega+\delta) \nonumber \\
&=& \frac{1}{M}\sum_{l=1}^{M} \lim_{\delta\downarrow 0} \sum_{ \epsilon_\alpha^{(l)} \in \left[ \omega-\delta ,  \omega+\delta \right] } \Psi_\alpha^{(X)(l)} (r_j)   \Psi_\alpha ^{(X)(l)}(r_{j^\prime } )\nonumber. \\
\label{eq:greensFctImaginaryPart}
\end{eqnarray}
In the wave vector representation, the Green's function can be evaluated using
\begin{equation}
\overline G_{k,k^\prime}^{(X)} (z) = \frac{1}{M}\sum_{l=1}^{M} \sum_{\alpha }  \frac{ \Psi_\alpha^{(X)(l)} (q_k)   \Psi_\alpha^{(X)(l)*}(q_{k^\prime } ) }{z+i\epsilon_\alpha^{(l)} }.
\label{eq:greensFktMomentum}
\end{equation}
As the system is translational invariant in a stochastic sense, the disorder-averaged Green's function is diagonal in the wave vector basis, i.e., $ \overline G_{k,k^\prime}^{(X)} (z)  \propto \delta_{k,k^\prime }$ .

\subsection{Asymptotic behavior in position space}

As  the disorder-averaged Green's function is  translationally invariant, it can be  expressed as the difference of  two positions, i.e., 
\begin{equation}
\eta_{X, r} (\omega) \equiv \lim_{N\rightarrow \infty} \eta_{j,j^\prime}^{(X)} (\omega),
\label{eq:greenFctImaginaryPartSeparation}
\end{equation}
where $ r=\left| r_j -r_{j^\prime }\right| $. Importantly, the asymptotic behavior  of $ \eta_{X,r}(\omega )$ for large $r $  exhibits the same scaling as  Eq.~\eqref{eq:greensFktAsymptoticBehavior}.

In Fig.~\ref{figComparisonGreensFctWaveFct}, we depict $  \eta_{X,r}^{2} (\omega)/\mathcal N$ for both $X=\text{L},\text{M}$ as a function of position by a solid green line. The normalization $\mathcal N$ is chosen such that the integral over the position is one.  We depicted the squared function to allow for a better comparison with the disorder-averaged eigenstates defined later in Eq.~\eqref{eq:disordereAveragedProbabilities}.

The decay of the Green's function with  increasing separation $ \left| r\right|$ is clearly visible for both photon and molecule Green's functions. For comparison, we have added the exponential decay predicted by the analytical coherence length $\zeta_{\text{coh},1}$ in Eq.~\eqref{eq:coherenceLength} as a dash-dotted black line.  We observe that the numerical and analytical calculations  precisely agree, which is especially clearly visible  for the molecule Green's function. 
We thus conclude that the asymptotic behavior of the Green's function is determined by $\zeta_{\text{coh}}=\zeta_{\text{coh},1}$ and that $a_1\gg a_2$ in Eq.~\eqref{eq:greensFktAsymptoticBehavior}. This assessment is additionally confirmed by the agreement of the analytical and numerical  calculations of the coherence length shown in Fig.~1 (g)-(i) in the Letter .

In Fig.~\ref{figComparisonGreensFctWaveFct}, the photon Green's function is more smooth than the matter contribution and exhibits a modulation as a function of separation $r$. The modulation frequency is determined by the real part of the root $q_{*,1}$ in Eq.~\eqref{eq:greensFktNonAnalyticities}. We explain the deviation from a mononchromatic modulation by the finite energy integral, over which we average the Green's function in Eq.~\eqref{eq:greensFctImaginaryPart}.

\subsection{Disorder-averaged eigenstates}
\label{eq:disorderAveragedEigenstate}

We define the disorder-averaged eigenstates in position space for the eigenstates $\Psi_\alpha^{(X)}(r) $ within a small energy interval $ \epsilon_\alpha \in \left[ \omega-\delta ,  \omega+\delta \right] $ as
\begin{equation}
\overline  {\left| \Psi^{(X)}(r)\right|^2 }=\frac{1}{N_{\omega,\delta}} \sum_{ \epsilon_\alpha \in \left[ \omega-\delta ,  \omega+\delta \right]}  \left|\Psi_\alpha^{(X)}(r-\overline{ r_\alpha})\right|^2  ,
\label{eq:disordereAveragedProbabilities}
\end{equation}  
where $x\in \left\lbrace \text{L}, \text{M}\right\rbrace$, $N_{\omega,\delta}$ is the number of eigenstates in the energy interval, and the center of each eigenstate $\alpha$ is given by
\begin{equation}
\overline{ r_\alpha} = \frac{ \sum_{j=1}^{N} r_j  \left|\Psi_\alpha^{(X)}(r_j)\right|^2  } {\sum_{j=1}^{N}  \left|\Psi_\alpha^{(X)}(r_j)\right|^2 }.
\label{eq:eigenStateCenter}
\end{equation}
Similarly, we can define the disorder-averaged eigenstates in wave vector space, which agrees with the Green's function in Eq.~\eqref{eq:greensFktMomentum} for $k=k^\prime$.

Guided by the analytical and numerical calculations of the Green's function, we anticipate that the exponential decay of the disorder-averaged wavefunction is dominated by one exponential term.
Similar to the Green's function in Eq.~\eqref{eq:greensFktAsymptoticBehavior}, we thus define the localization length $\zeta_{\text{loc}}$ such that 
\begin{equation}
\overline  {\left| \Psi^{(X)}(r)\right|^2 }   \propto \exp \left[ -\frac{\left| r\right| }  {\zeta_{\text{loc}}(\omega) }\right]
\end{equation}
for large $\left| r\right| $. 

In Fig.~\ref{figComparisonGreensFctWaveFct}, we depict the disorder-averaged eigenstates and observe that the photon contribution of the wave function exponentially decays with $\left| r \right|$. The fluctuations are significantly less than  the fluctuations in  the Green's function. Interestingly, we observe that the disorder-averaged eigenstates and the Green's function exhibit the same decay behavior for small-to-intermediate $\left| r \right|$, i.e., $\zeta_{\text{coh}} \approx \zeta_{\text{loc}}$. We attribute this  agreement to the photon-mediated long-range interaction between distant molecules, which protects their coherent phase relation.

For very large $\left| r \right|$, the eigenstates decay significantly slower than the Green's function. These deviations can have two distinct explanations: (i) the Green's function is subject to significant destructive interference because of the average over the finite energy interval in Eq.~\eqref{eq:greensFctImaginaryPart}. (ii) The numerical calculation suffers from a finite computational precision. Importantly, we have numerically checked that this slow decay is not determined  by the branch-cut related value $\zeta_{\text{coh},2}$ in Eq.~\eqref{eq:coherenceLength}, which would predict a significantly faster decay rate.

\subsection{Numerical calculation of the coherence length}

Because of the large fluctuations of the Green's function, extracting the coherence  length via fitting is numerically unstable. Instead, we determine the coherence length  via comparison with the position variance of the normalized  function as follows:

We start with a generic exponentially decaying  function 
\begin{eqnarray}
F(r) &=& \frac{\lambda}{2 } e^{-\lambda \left| r \right| } ,
\end{eqnarray}
whose integral over $x\in \left\lbrace -\infty, \infty\right\rbrace$ is one. In particular, we use $F(r)  =\eta_{\text{M},r}^2(\omega)/\mathcal N$. Here, we use $X=\text{M}$, as the molecule Green's function does not exhibit (almost) monochromatic smooth modulations like the photon Green's function (c.f., Fig.~\ref{figComparisonGreensFctWaveFct}). While for the  monochromatic modulation of the  photon Green's function, $F(r)$ must be multiplied by a cosine function with an unknown frequency, the high-frequent fluctuations of the molecule Green's function simply average away.
The  position variance of $F(r)$ evaluates to
\begin{eqnarray}
\text{Var}_{F} \hat X  & = & 2\int_{0}^{\infty}\frac{\lambda}{2 }  r^2   e^{-\lambda r }dr  \nonumber \\
& = &  \lambda \frac{1}{\lambda^2}\int_{0}^{\infty}   e^{-\lambda x }dr \nonumber \\
& = &  \lambda \frac{1}{\lambda^2}\frac{1}{\lambda } \nonumber \\
& = &  2 \lambda\frac{1}{\lambda^3 }   =  2 \frac{1}{\lambda^2 }  = 2 \zeta_{\text{coh}}^2 .
\end{eqnarray}
In the last step, we have inserted the coherence length $\zeta_{\text{coh}} = 1/\lambda$. 

This numerical approach assumes that there is only one exponentially decaying term, as opposed to the two terms predicted in Eq.~\eqref{eq:greensFktAsymptoticBehavior}. However, the analysis in Fig.~\ref{figComparisonGreensFctWaveFct} has shown that only the pole of the Green's function substantially determines the decaying behavior, while the branch-cut term can be neglected. For this reason, the deployed numerical procedure provides a value for the coherence length related to the  Green's function pole.
We emphasize that the precise agreement of the analytical and numerical calculations in Fig. 1(g)-(i) in the Letter justifies the numerical approach introduced here.

\end{document}